\documentclass[twocolumn,nofootinbib,prd,aps,superscriptaddress,tightenlines]{revtex4-1}

\usepackage{slashed}
\usepackage{graphicx}
\usepackage{pstricks}
\usepackage{braket}
\usepackage{bm}

{\count255=\time\divide\count255 by 60 \xdef\hourmin{\number\count255}
  \multiply\count255 by-60\advance\count255 by\time
  \xdef\hourmin{\hourmin:\ifnum\count255<10 0\fi\the\count255}}

\newcommand{\LL}{\mathsf{L}}
\newcommand{\e}{\epsilon}
\newcommand{\nn}{\nonumber \\ }

\newcommand{\Ls}{\mathsf{L}_{s}}
\newcommand{\Lt}{\mathsf{L}_{t}}
\newcommand{\Lu}{\mathsf{L}_{u}}
\newcommand{\co}{\,,}
\newcommand{\fs}{\, .}

\newcommand{\ga}{\mathcal{G}_f}
\newcommand{\Rc}{\bar{R}}

\newcommand{\X}{\Lambda_Q}

\newcommand{\al}{\rho}

\def\GAC{\mathcal{G}_+}
\def\GFF{\mathcal{G}_{TT}}

\def\trcs#1{ \mathcal{T}^{(CS)}_{#1}}
\def\trwf#1{ \mathcal{T}^{(WF)}_{#1}}
\def\ssab{\Phi_B}
\def\ssba{\Phi_C}
\def\ssi{ \Phi_1}
\def\ssii{\Phi_2}
\def\sscs{\Phi_{CS}}
\def\sswf{\Phi_{WF}}
\def\lamQ{\Lambda_{Q}}
\def\lamphi{\Lambda_{\phi}}
\def\mphi{\mathcal{M}_\phi}

\begin{document}

\title{Radiative Corrections to Longitudinal and Transverse Gauge Boson and Higgs Production}

\author{Andreas Fuhrer}


\author{Aneesh V.~Manohar}

\affiliation{Department of Physics, University of California at San Diego,
  La Jolla, CA 92093}

\author{Jui-yu Chiu}

\affiliation{Department of Physics, Carnegie Mellon University,
  Pittsburgh, PA 15213}

\author{Randall Kelley}

\affiliation{Department of Physics, Harvard University, Cambridge, MA
  02138}

\begin{abstract}
Radiative corrections to gauge boson and Higgs production computed recently using soft-collinear effective theory (SCET) require the one-loop high-scale matching coefficients in the standard model. We give explicit expressions for the matching coefficients for the effective field theory (EFT) operators for $q \bar q \to VV $ and $q \bar q \to \phi^\dagger \phi$ for a general gauge theory with an arbitrary number of gauge groups. The group theory factors are given explicitly for the standard model, including both QCD and electroweak corrections. 
\end{abstract}


\maketitle

\section{Introduction}

QCD and electroweak radiative corrections to high energy scattering amplitudes were computed recently using effective field theory methods~\cite{Chiu:2009ft,Chiu:2009mg,Chiu:2009yz,Chiu:2009yx,Chiu:2008vv,Chiu:2007dg,Chiu:2007yn}, by extending SCET~\cite{BFL,SCET1,SCET2,BPS} to broken gauge theories with massive gauge bosons. The radiative corrections (including the purely electroweak ones) are large because of Sudakov double-logarithms; for example the electroweak corrections to transverse $W$ pair production are 37\% at 2~TeV. The computation of radiative corrections is divided into a matching computation from the standard model onto SCET at a high scale $Q$ of order the center-of-mass energy $\sqrt{s}$, and the scattering amplitude in the effective theory. Logarithms of the form $\log^2 Q^2/M_Z^2$, including the Sudakov double logarithms, are summed using renormalization group evolution in the effective theory. The high-scale matching coefficients for vector boson and Higgs production were included in the numerical results of Refs.~\cite{Chiu:2009ft,Chiu:2009mg}. In this paper we give a detailed discussion of the required matching calculation, and explicit results for a gauge theory with an arbitrary number of gauge groups, as well as results for the $SU(3)\times SU(2) \times U(1)$ standard model theory.

The computation of radiative corrections to gauge boson and Higgs production is not new, and has been obtained previously by fixed order calculations by many groups~\cite{ccc,ciafaloni1,ciafaloni2,fadin,kps,fkps,jkps,jkps4,beccaria,dp1,dp2,hori,beenakker,dmp,pozzorini,js,melles1,melles2,melles3,Denner:2006jr,kuhnW,Denner:2008yn,Ellis:1985er,Kunszt:1993sd,sack,bardin,Dixon:1998py,Dixon:1999di}. However, the matching computation we need is not readily available in the literature. What is available is the total one-loop scattering amplitude, which is the sum of the matching coefficient and the SCET amplitude, and our results agree with existing computations for the sum. The EFT computation requires the matching and SCET contributions separately, so that large logarithms can be summed using renormalization group evolution in the effective theory.

In Sec.~\ref{sec:smatrix}, we show how the matching computation is related to the $S$-matrix for parton scattering. Using this, we can use the matching coefficients to compute the one-loop corrections to the $q \bar q \to g g$ cross-section in QCD. This was computed a long time ago by Ellis and Sexton~\cite{Ellis:1985er}, and we have checked that our amplitude reproduces their cross-section. Kunszt, Signer and Tr\'ocs\'anyi~\cite{Kunszt:1993sd} give the helicity amplitudes for $q \bar q \to gg$ for an $SU(N)$ gauge theory, and we agree with their results.

In this paper, we give expressions for the one-loop matching contributions for gauge boson pair production and scalar production. These can then be used to compute the renormalization group improved scattering amplitudes for transverse and longitudinal gauge boson pair production, as well as Higgs production, using the results in Refs.~\cite{Chiu:2009ft,Chiu:2009mg}. We give the results for the individual Feynman diagrams as the product of a Feynman integral and a group theory factor. The results can be used for an arbitrary gauge theory with any number of gauge groups. Gauge bosons from a maximum of three gauge groups can occur in a single diagram at one loop.

Section~\ref{sec:outline} gives an outline of the method we use to compute the matching condition. We discuss the relation between the on-shell diagrams in dimensional regularization and the matching calculation, the group theory notation, kinematics and the Dirac basis for the matrix elements. The diagrams for vector boson production are given in Sec.~\ref{sec:vector}, and the standard model amplitude is given in Sec.~\ref{sec:smvv}. Section~\ref{sec:scalar} gives the graphs for scalar production, with the standard model results, including top-quark loops, in Sec.~\ref{sec:smscalar}.
A consistency check between the matching condition and the EFT anomalous dimension matrix is verified in Sec.~\ref{sec:check}. Section~\ref{sec:smatrix} gives the relation between the matching calculation and the on-shell $S$-matrix elements in the massless theory.

\section{Outline of Method and Notation}\label{sec:outline}

The basic processes we consider are $f(p_1) + \bar{f}(p_2) \to V_i^a(p_4) + V_j^b(p_3)$ and $f(p_1) + \bar{f}(p_2) \to \phi^\dagger(p_4) + \phi(p_3)$. Here $f$ and $\bar f$ are incoming fermions and antifermions of momentum $p_1$ and $p_2$, $V^a_i(p)$ is a gauge boson of gauge group $G_i$ with gauge index $a$ and momentum $p$, and $\phi$ is a (complex) scalar field. Note that $i$ and $j$ can refer to different gauge groups, so that our results are also applicable to processes such as $q \bar q \to W g$. The gauge bosons $V$ will be taken to have transverse polarization. Massive gauge bosons which are longitudinally polarized can be computed using the $\phi^\dagger \phi$ amplitude and the Goldstone boson equivalence theorem.

Our EFT results are valid in the regime where $\sqrt{s}$ is much larger than the gauge boson masses $M_{W,Z}$, the Higgs mass $M_Z$, and the fermion masses. The matching from the full gauge theory onto the EFT is done at a scale $\mu$ of order $\sqrt{s}$, and power corrections such as $M_Z^2/s$ are neglected. Thus the matching coefficients can be computed by evaluating the graphs in the full theory setting all the particle masses to zero, and neglecting gauge symmetry breaking. For the standard model, this implies that the best way to compute the EFT operators is to match onto operators with $W^{1,2,3}$ and $B$ fields, rather onto operators with $W^{\pm}, Z$ and $A$ fields.

We first summarize the standard method used to evaluate matching conditions for an EFT. More details can be found, for example, in Refs.~\cite{Manohar:1996cq,Manohar:1997qy}. The full theory graphs are evaluated using dimensional regularization in $4-2\epsilon$ dimension, which regulate the ultraviolet (UV) and infrared (IR) divergences, and have the schematic form
\begin{eqnarray}
A_{\text{full}} &=& \left(\sum_{k\ge1} \frac{C_k}{\epsilon^k}\right)_{\text{UV}}+\left(\sum_{k\ge1} \frac{D_k}{\epsilon^k}\right)_{\text{IR}}+A_{\text{full,finite}}\,.
\end{eqnarray}
The ultraviolet divergences are cancelled by the full theory renormalization counterterms, leaving the infrared divergences,
\begin{eqnarray}
A_{\text{full}} + \text{c.t.} &=& \left(\sum_{k\ge1} \frac{D_k}{\epsilon^k}\right)_{\text{IR}}+A_{\text{full,finite}}\,.
\label{afull}
\end{eqnarray}
The EFT graphs are also computed using dimensional regularization. Since all the scales that enter the EFT computation (such as masses) have been set to zero, the EFT integrals are all scaleless and vanish.
The EFT integrals have the schematic form
\begin{eqnarray}
A_{\text{EFT}} &=& \left(\sum_{k\ge1} \frac{\widetilde C_k}{\epsilon^k}\right)_{\text{UV}}+\left(\sum_{k\ge1} -\frac{\widetilde C_k}{\epsilon^k}\right)_{\text{IR}}=0\,,\end{eqnarray}
i.e.\ a cancellation of $1/\epsilon$ terms arising from ultraviolet and infrared divergences, \emph{without any finite part}. The
$(1/\epsilon)_{\text{UV}}$ terms are cancelled by the renormalization counterterms in the EFT, leaving the $(1/\epsilon)_{\text{IR}}$ terms,
\begin{eqnarray}
A_{\text{EFT}}  + \text{c.t.} &=& \left(\sum_{k\ge1} -\frac{\widetilde C_k}{\epsilon^k}\right)_{\text{IR}}\,,
\label{aeft}
\end{eqnarray}
The counterterms (and hence the anomalous dimensions) in the full and effective theories are in general different.
The EFT, by construction, is designed to reproduce the infrared structure of the full theory. Thus the $(1/\epsilon)_{\text{IR}}$ in the full and effective theories \emph{must} agree, 
\begin{eqnarray}
D_k= -\widetilde C_k\,,
\label{eq5}
\end{eqnarray} 
which provides a non-trivial computational check on the EFT, and also shows that infrared divergences in the full theory are equal to ultraviolet divergences in the EFT.

The matching coefficient is given by the difference of the renormalized full and effective theory expressions, Eqs.~(\ref{afull},\ref{aeft}). Using Eq.~(\ref{eq5}), we see that the matching coefficient is $A_{\text{full,finite}}$. This gives the standard method of computing matching coefficients --- compute graphs in the full theory in dimensional regularization setting all EFT scales to zero, and keep only the finite parts. This is the procedure used here. In giving the values for the graphs, we will also give the divergent terms, which should be dropped for the matching corrections. The divergent terms are useful in that they allow one to check the matching of infrared divergences, and also to compare with the results of Refs.~\cite{Ellis:1985er,Kunszt:1993sd}. Scaleless integrals in the full theory computation have been set to zero, so the $1/\epsilon$ divergences can be either UV or IR.

\subsection{Group Theory}

We consider an arbitrary gauge group $\otimes_r G_r$ which is a product
of groups with coupling constants $g_r=\sqrt{4 \pi \alpha_r}$.
The generators  of $G_r$ are $T^a_r$ and  satisfy the commutation relations
\begin{equation}
\left[T^a_r,T^b_s\right] = i f^{(r)}_{abc}  \delta_{rs}\, T^c_r\,,
\end{equation}
where $f^{(r)}_{abc}$ are the structure constants of $G_r$. Some products of group generators can be simplifed in terms of Casimir operators, e.g.\
\begin{eqnarray}
T_j^b T^a_i T^b_j &=& \left(C_R(j) - \frac12 \delta_{ij}
C_A(i)\right)T^a_i \co
\end{eqnarray}
where $C_R$ is the Casimir of the representation $R$ of the matrices $T_j$, and $C_A(i)$ is the Casimir of the adjoint representation of $G_i$.

In general, anti-commutators of group generators such as $\left\{T^a_r,T^b_r\right\}$ cannot be simplified. If $G_r$ is an $SU(N_r)$ group, and $T_r$ is in the fundamental representation, one has 
\begin{equation}
\left\{T^a_r,T^b_r\right\} = \frac{1}{N_r}\delta_{ab}^{(r)}+d^{(r)}_{abc} T^c_r \co
\label{eq8}
\end{equation}
where $d_{abc}=0$ for $SU(2)$. However, there is no simple expression such as Eq.~(\ref{eq8}) in general, not even for arbitrary representations of $SU(N)$. For this reason, we will give a general expression for the group theory factor valid for arbitrary gauge theories, and then its value for a $SU(N) \times SU(2) \times U(1)$ gauge theory.

Diagrams with a closed fermion or scalar loop contribute at one loop order. We use the symbols $\text{Tr}_{WF}$ and  $\text{Tr}_{CS}$  to denote traces over the Weyl fermions and the complex scalars of the theory, respectively.

For the standard model results, $T^a$ are the color generators, $t^a$ are the $SU(2)$ generators, and $Y$ is the $U(1)$ generator.

\subsection{Kinematics}

The amplitude $\mathcal{M}$ is defined as
\begin{equation}
\langle p_3\,p_4\,,\mathrm{out}| p_1\,p_2\,,\mathrm{in} \rangle =
 (2\pi)^4 \delta^{(4)}(p_1+p_2-p_3-p_4)i \mathcal{M} \fs \nonumber
\end{equation}
We will work in the center of mass frame (CMS) throughout this
article. For $f(p_1) + \bar{f}(p_2) \to V_i^a(p_4) +
V_j^b(p_3)$, the Dirac structure can be written as a linear combination of five
basic terms
\begin{eqnarray}
\mathcal{M}_0 &=& \bar{v}(p_2) \slashed{\epsilon}_4
\left(\slashed{p}_4-\slashed{p}_2\right) \slashed{\epsilon}_3 P_L
u(p_1) \co\nn
\mathcal{M}_1 &=& \bar{v}(p_2) \slashed{p}_4 (\epsilon_4 \cdot
\epsilon_3) P_L u(p_1) \co\nn
\mathcal{M}_4 &=& \bar{v}(p_2) \slashed{\epsilon}_4 (\epsilon_3 \cdot
p_1) P_L u(p_1) \co\nn
\mathcal{M}_5 &=& - \bar{v}(p_2) \slashed{\epsilon}_3 (\epsilon_4
\cdot p_2)P_L u(p_1) \co\nn
\mathcal{M}_6 &=& \bar{v}(p_2) \slashed{p}_4 (\epsilon_4 \cdot p_2)(
\epsilon_3 \cdot p_1) P_L u(p_1) \co
\end{eqnarray}
in the notation of Sack~\cite{sack}, where $\epsilon^\mu_i \equiv \epsilon^\mu(p_i)$ and $P_{L} \equiv \left(1-\gamma_5 \right)/2$.  The other amplitudes used by Sack ($\mathcal{M}_{2,3,7,8,9}$) vanish for \emph{transversely} polarized \emph{on-shell} gauge bosons, neglecting power corrections.

For scalar production, the Dirac structure which enters is
\begin{eqnarray}
\mphi &=& \bar{v}(p_2) \slashed{p}_4 P_L
u(p_1) \fs
\end{eqnarray}

The full amplitude is the sum of all diagrams $R_i$ with group theory factor $\mathcal{C}_i$,
\begin{equation}
\mathcal{M} = \sum_{i} \mathcal{C}(R_i) R_i \fs
\end{equation}
In many cases, a diagram $R$ has a corresponding crossed graph which we denote by
$\Rc$ with group theory factor $\mathcal{C}(\Rc)$.

The Mandelstam variables are defined as 
\begin{eqnarray}
s &=& (p_1+p_2)^2 \co\nn
t &=& (p_1-p_4)^2 \co \nn
u &=& (p_1-p_3)^2 \fs
\end{eqnarray}
to agree with the conventions of Refs.~\cite{Chiu:2009ft,Chiu:2009mg}.

Under the exchange of the two final state gauge bosons, $\epsilon_3
\leftrightarrow \epsilon_4$, $p_3 \leftrightarrow p_4$, the matrix
elements and Mandelstam variables transform as
\begin{eqnarray}
\mathcal{M}_0 &\leftrightarrow& \mathcal{M}_0+2\mathcal{M}_1\co\nn
\mathcal{M}_1 &\rightarrow& -\mathcal{M}_1\co\nn
\mathcal{M}_4 &\leftrightarrow& \mathcal{M}_5 \co \nn
\mathcal{M}_6 &\leftrightarrow& -\mathcal{M}_6 \co \nn
t &\leftrightarrow& u \co \nn
s &\leftrightarrow & s \fs
\label{mcross}
\end{eqnarray}
If there is a crossed graph, then $\Rc$ is obtained from $R$ using Eq.~(\ref{mcross}).

Throughout the article, space-time is $d = 4-2\epsilon$
dimensional which regulates the ultraviolet as well as the infrared
behavior, and we work in 't~Hooft-Feynman gauge, $\xi = 1$. Furthermore, we define the function $\mathsf{L}_X \equiv
\log(-X-i0^+)/\mu^2$. For scattering kinematics, $s > 0$ and $t,u < 0$, the correct
analytical continuation is given by
\begin{eqnarray}
\Ls &=& \log(s/\mu^2)-i\pi\co\nn \Lt &=& \log(-t/\mu^2)\co \nn \Lu &=&
\log(-u/\mu^2) \fs
\end{eqnarray}

We have assumed that the incoming fermion is a left-chiral field, so that the incoming fermion $f$ has helicity $h=-1/2$ and incoming antifermion $\bar f$ has helicity $h=1/2$. The results for a right-chiral field are given by $P_L \to P_R$. 

\subsection{EFT Lagrangian}

We give the Feynman diagram results for the on-shell scattering amplitude $\mathcal{M}$. This also gives the matching condition onto the SCET operators in the EFT. The EFT Lagrangian is
\begin{eqnarray}
L &=& \frac12\sum_{p_1,p_2,p_3,p_4} \mathcal{M}^{ia,jb}(p_1,p_2,p_3,p_4) V^{i,a}_{p_4} V^{j,b}_{p_3}
\bar \psi_{p_2} \psi_{p_1}\nn
\end{eqnarray}
for vector boson production, and 
\begin{eqnarray}
L &=& \sum_{p_1,p_2,p_3,p_4} \mathcal{M}^{ia,jb}(p_1,p_2,p_3,p_4) \phi^\dagger_{p_4} \phi_{p_3}
\bar \psi_{p_2} \psi_{p_1}\nn
\end{eqnarray}
for scalar production. The subscripts $p_i$ are the label momenta of the external SCET fields, and are summed over. 

The vector boson term has a factor of $1/2$ because there are two identical fields. To make clear the combinatorial factor of $1/2$, consider the production of a $W$ boson with momentum $p_W$ and a gluon with momentum $p_g$. This is obtained from $\mathcal{M}$ by picking out the term with $i,a$ in $SU(2)$ and $j,b$ in $SU(3)$, and setting $p_4=p_W$ and $p_3=p_g$ \emph{or} the term with $i,a$ in $SU(3)$ and $j,b$ in $SU(2)$, and setting $p_4=p_g$ and $p_3=p_W$, \emph{but not both.}

\subsection{Topologies}

\begin{figure}
\begin{center}
\begin{tabular}{ccc}
\includegraphics[height=1.5cm]{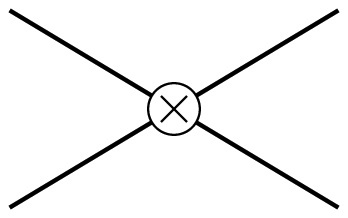} &
\includegraphics[height=1.5cm]{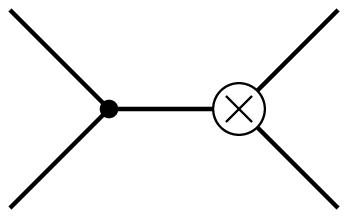} &
\includegraphics[height=1.5cm]{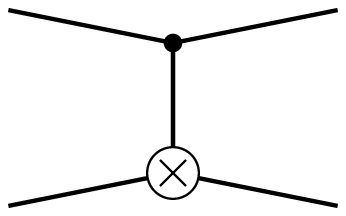} \\
T1 & T2 & T3 \\[10pt]
\includegraphics[height=1.5cm]{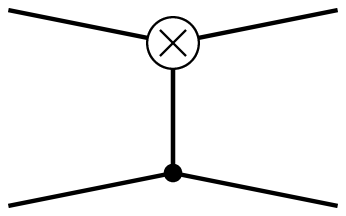} &
\includegraphics[height=1.5cm]{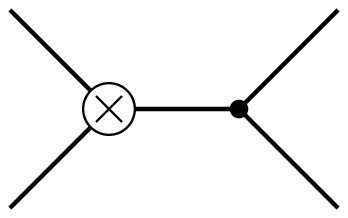} &
\includegraphics[height=1.5cm]{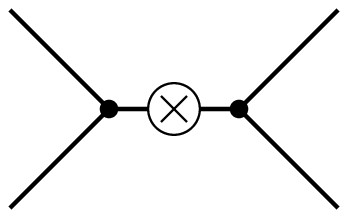} \\
T4 & T5 & T6 \\[10pt]
\includegraphics[height=1.5cm]{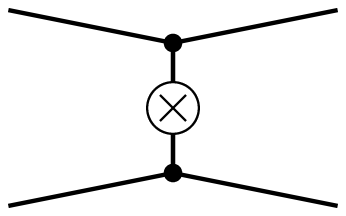} & & \\
T7 &  &  
\end{tabular}
\end{center}
\caption{The seven different topologies for a general $2 \to 2$
  scattering process. The $\otimes$ denotes a one particle irreducible
  subdiagram. Wavefunction renormalization diagrams are omitted.}\label{fig:topos}
\end{figure}

The diagrams are classified in seven different topologies shown in
Figure \ref{fig:topos}. Note that we do not explicitly draw the crossed
topologies. Because this is a matching calculation, counterterm diagrams and wavefunction corrections are omitted. The on-shell wavefunction graphs are scaleless, and vanish in dimensional regularization.

\section{Diagrams for vector boson production}\label{sec:vector}

We provide the result of each tree-level and one-loop diagram $R_i$ and list the group theory
structure $\mathcal{C}_i$ in a general form in terms of generators of the gauge
groups. The pertinent group theory factors for the Standard Model are given in the Section~\ref{sec:smvv}.

\subsection{Tree level amplitude}

\begin{figure}[tb]
\begin{center}
\begin{tabular}{cc}
\includegraphics[height=1.5cm]{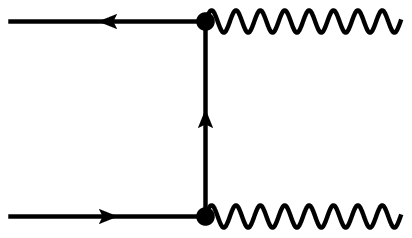}&\includegraphics[height=1.5cm]{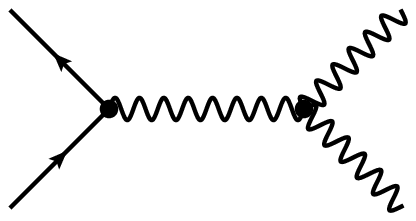}
\\
$R_1$ & $R_2$ 
\end{tabular}
\end{center}
\caption{The tree level diagrams. Quarks, gauge bosons, scalars and ghosts are denoted by solid, wavy, dashed and dotted lines, respectively. Crossed diagrams are not shown.}\label{fig:tree}
\end{figure}

The tree level diagrams are shown in Figure~\ref{fig:tree}.
For the tree level amplitude, the group theory factors and the
diagrams read
\begin{eqnarray}
\mathcal{C}(R_1) &=& g_i g_j T^b_j T^a_i \co \nn
\mathcal{C}(\Rc_1) &=& g_i g_j  T^a_i T^b_j \co \nn
\mathcal{C}(R_2) &=& g_i^2 \left(-i \delta_{ij} f^{(i)}_{abc}T_i^c\right)
\end{eqnarray}
\begin{eqnarray}
R_1 &=& -\frac{1}{t}\left(\mathcal{M}_0 +2 \mathcal{M}_1\right) \co \nn
\Rc_1 &=& -\frac{1}{u}\left(\mathcal{M}_0\right)\co  \nn
R_2 &=& -\frac{1}{s}\left(2 \mathcal{M}_1\right) \fs
\end{eqnarray}
where $\Rc_1$ is the crossed-graph related to $R_1$. $R_2$ does not have a crossed graph.

\begin{widetext}

\subsection{Topology T1}

The four diagrams shown in Figure \ref{fig:t1} share topology T1. 

\begin{figure}
\begin{center}
\begin{tabular}{cc}
\includegraphics[height=1.5cm]{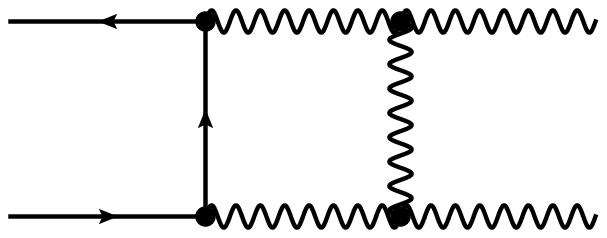}&\includegraphics[height=1.5cm]{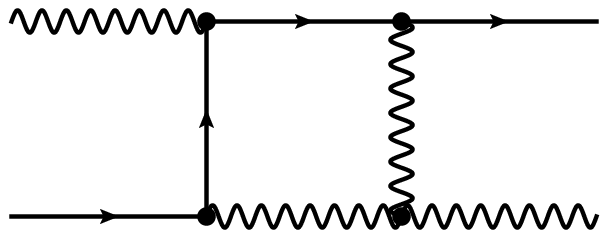}
\\
$T1a$ & $T1b$ \\[10pt]
\includegraphics[height=1.5cm]{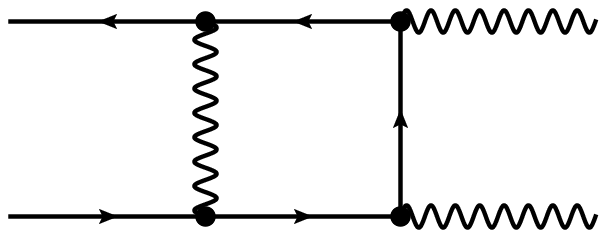} &
\includegraphics[height=1.5cm]{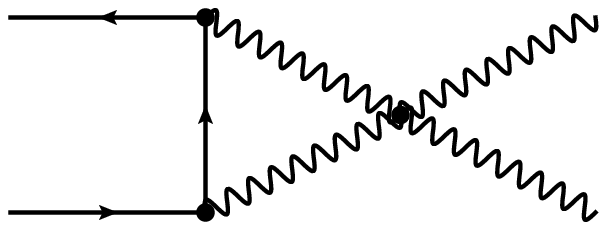} \\
$T1c$ & $T1d$ 
\end{tabular}
\end{center}
\caption{Diagrams with topology T1. See caption of Figure~\ref{fig:tree}.}\label{fig:t1}
\end{figure}

\subsubsection{T1a}

\begin{equation}
\mathcal{C}(R_{T1a}) = \frac{g_i^4}{16\pi^2}\delta_{ij} f_{ebc}^{(i)} f_{aed}^{(i)} T^c_i T^d_i
\end{equation}

\begin{eqnarray}
R_{T1a} &=& \frac{\mathcal{M}_0}{t}\Bigg\{-\frac{2}{\epsilon^2 }+\frac{2 (\Ls-1)}{\epsilon } + \frac{1}{u}\biggl[
-3t\Ls^2-(s+4t)\Lt^2+2(s+4t)\Ls \Lt +2u \Lt -\pi^2\left(\frac76s+\frac{25}{6}t\right)-4u\biggr]\Biggr\}\nn
&&+ \mathcal{M}_1\Biggl\{-\frac{1}{\epsilon^2 }\left(\frac9s+\frac4t\right)+\frac{1}{\epsilon } 
\left(\frac{4\Ls}{t}+\frac{8\Lt}{s}+\frac{\Ls}{s}-\frac{2}{s}-\frac{4}{t}\right)+ \frac{1}{u^2ts}\biggl[\frac12t(9s^2+14st+7t^2)\Ls^2+s(2s+t)(s+2t)\Lt^2\nn
&&-2(2s^3+9s^2t+10st^2+4t^3)\Ls\Lt-2t^2u\Ls-2us(2s+3t)\Lt+\pi^2\Bigl(\frac73s^3+\frac{125}{12}s^2t+\frac{71}{6}st^2+\frac{19}{4}t^3\Bigr)\nn
&&-8s^3-20s^2t-16st^2-4t^3
\biggr]\Biggr\}\nn
&&+ \frac{\mathcal{M}_4+\mathcal{M}_5}{t}\Biggl\{-\frac{2}{\epsilon^2}+\frac{1}{\epsilon
}\left(\frac{2t}s+2\Lt\right)+ \frac{1}{u^2}\biggl[-t(3s+4t)\Ls^2 -(s^2+5st+5t^2)\Lt^2+2t(3s+4t)\Ls\Lt \nn
&&+2ut(2s+t)\frac{\Ls}{s}- 2ut\Lt  +\pi^2\Bigl(\frac{s^2}6-\frac83st-\frac{23}{6}t^2\Bigr) +4\frac{t^3}{s}+4st+8t^2\biggr]\Biggr\}\nn
&&+ \frac{\mathcal{M}_6}{tu^3}\Biggl\{-4t(s+2t)(\Ls-\Lt)^2 +4u(3s+5t)(\Ls-\Lt)-4\pi^2t\Bigl(s+2t\Bigr)-4u^2 \Biggr\}
\end{eqnarray}

The crossed graph $\Rc_{T1a}$ is given by applying Eq.~(\ref{mcross}) to $R_{T1a}$, and has color factor
\begin{equation}
\mathcal{C}(\bar  R_{T1a}) = \frac{g_i^4}{16\pi^2} \delta_{ij} f_{eac}^{(i)} f_{bed}^{(i)} T^c_i T^d_i
\end{equation}
given from $\mathcal{C}(R_{T1a})$ by $i,a \leftrightarrow j,b$.

\subsubsection{T1b}

\begin{eqnarray}
\mathcal{C}(R_{T1b}) &=& \frac{g_i^3 g_j}{16\pi^2}  (if_{dac}^{(i)}) T^c_i T^b_j T^d_i
\end{eqnarray}
The diagram is given by
\begin{eqnarray}
R_{T1b}&=& \frac{\mathcal{M}_0}{tu}\Biggl\{\frac{2s}{\epsilon^2 }+\frac{2u \Lu +2t \Lt+s}{\epsilon }
+\frac{1}{s^2}\biggl[-st(2s+3t)\Lu^2
+su \left(s+3t\right)\Lt^2 +2s(s^2+3st+3t^2)\Lu\Lt +s^2t\Lu+s^2u\Lt\nn
&&-\pi^2\left(\frac76s^3+3s^2t+3st^2\right)+2s^3
\biggr] \Biggr\}\nn
&&+\frac{\mathcal{M}_1}{t}\Biggl\{-\frac{4}{\epsilon^2 }+\frac{4 \Lu-2}{\epsilon } + \frac{1}{s^2  u}\biggl[3stu\Lu^2 +su\left(2s+3t\right)\Lt^2 -2su\left(2s+3t\right)\Lu\Lt+2s^2u\Lt -\pi^2\left(\frac73s^3+\frac{16}{3}s^2t+3st^2\right)\nn
&&+4s^3+4s^2t
\biggr]\Biggr\}\nn
&&+ \frac{\mathcal{M}_4}{t u}\Biggl\{\frac{4s}{\epsilon }+
\frac{1}{s^2}\biggl[-3stu \left(\Lu-\Lt\right)^2 +4s^2t\Lu +4s^2u\Lt +\pi^2\left(3s^2t+3st^2\right) +8s^3\biggr]\Biggr\}\nn
&+& \frac{\mathcal{M}_5}{t u}\Biggl\{\frac{2s}{\epsilon^2}+\frac{2 u
  \Lt + 2 t \Lu}{\epsilon }+ \frac{1}{s^2}\biggl[st(2s+3t)\Lu^2-su(s+3t)\Lt^2 +6stu\Lu\Lt +\pi^2\Bigl(-\frac16s^3+3s^2t+3st^2\Bigr)\biggr]\Biggr\}\nn
&&+ \mathcal{M}_6\frac{12}{tu}(\Lt-\Lu) \fs
\end{eqnarray}

The crossed graph $\bar R_{T1b}$ is given by applying Eq.~(\ref{mcross}) to $R_{T1b}$, and has color factor
\begin{equation}
\mathcal{C}(\bar R_{T1b}) = \frac{g_i g_j^3}{16\pi^2}  (if_{dbc}^{(j)}) T^c_j T^a_i T^d_j
\end{equation}
given from $\mathcal{C}(R_{T1a})$ by $i,a \leftrightarrow j,b$.

\subsubsection{T1c}
\begin{eqnarray}
\mathcal{C}(R_{T1c})  &=& \sum_k \frac{g_i g_j g_k^2}{16\pi^2}  T^c_k T^b_j T^a_i T^c_k
\end{eqnarray}
\begin{eqnarray}
R_{T1c} &=& \frac{\mathcal{M}_0}{t}\left[\frac{2}{\epsilon^2 }-\frac{2 \Ls}{\epsilon } + 2 \Ls \Lt-\frac{7\pi^2}{6}-\Lt^2\right]\nn
&&+\frac{\mathcal{M}_1}{t}\Biggl\{\frac{4}{\epsilon^2 }-\frac{4 \Ls}{\epsilon } + \frac{1}{u^2}\biggl[st \Ls^2-(2s^2+3st+2t^2)\Lt^2+2(2s^2+3st+2t^2)\Ls\Lt+2tu(\Ls-\Lt)-\pi^2\Bigl(\frac73s^2+\frac{11}{3}st+\frac73t^2\Bigr)\biggr]\Biggr\}\nn
&&+\frac{\mathcal{M}_4+\mathcal{M}_5}{t}\Biggl\{\frac{4}{\epsilon }+ \frac{1}{u^2}\biggl[t(3s+2t) (\Ls-\Lt)^2+2ut\Ls +2u(2s+t)\Lt +\pi^2t\Bigl(3s+2t\Bigr) +8u^2\biggr]\Biggr\}\nn
&&+ \frac{\mathcal{M}_6}{tu^3}\Biggl\{4t(2s+t)(\Ls-\Lt)^2-4u(3s+t)(\Ls-\Lt)+4\pi^2t\Bigl(2s+t\Bigr)-4u^2\Biggr\}
\end{eqnarray}

The crossed graph $\bar R_{T1c}$ is given by applying Eq.~(\ref{mcross}) to $R_{T1c}$, and has color factor
\begin{equation}
\mathcal{C}(\bar R_{T1c}) = \sum_k \frac{g_i g_j g_k^2}{16\pi^2}  T^c_k T^a_i T^b_j T^c_k
\end{equation}
given from $\mathcal{C}(R_{T1c})$ by $i,a \leftrightarrow j,b$.

\end{widetext}

\subsubsection{T1d}
\begin{eqnarray}
\mathcal{C}(R_{T1d}) = \frac{g_i^4}{16\pi^2}\delta_{ij} \left\{T^d_i, T^c_i\right\}
f_{adg}^{(i)}f_{bcg}^{(i)} 
\end{eqnarray}

The result of the diagram is 
\begin{eqnarray}
R_{T1d}= \frac{\mathcal{M}_4+\mathcal{M}_5}{s}\left[
  \frac{2}{\epsilon}-2\Ls+4 \right] \fs
\end{eqnarray}

There is no corresponding crossed graph.

\subsection{Topology T2}

\begin{figure}
\begin{center}
\begin{tabular}{cc}
\includegraphics[height=1.5cm]{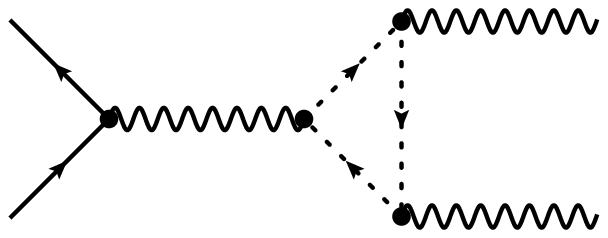}&\includegraphics[height=1.5cm]{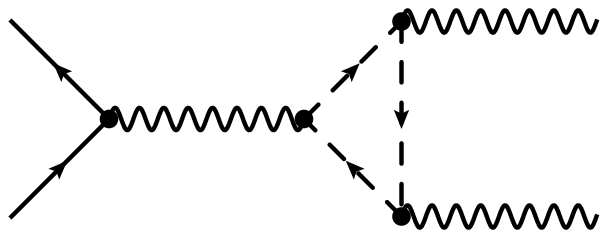}
\\
$T2a$ & $T2b$ \\[10pt]
\includegraphics[height=1.5cm]{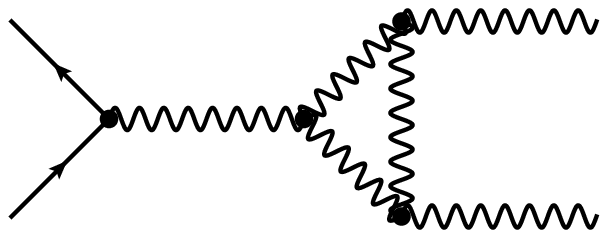} &
\includegraphics[height=1.5cm]{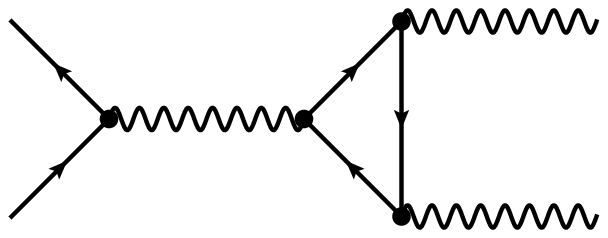} \\
$T2c$ & $T2d$ \\[10pt]
\includegraphics[height=1.5cm]{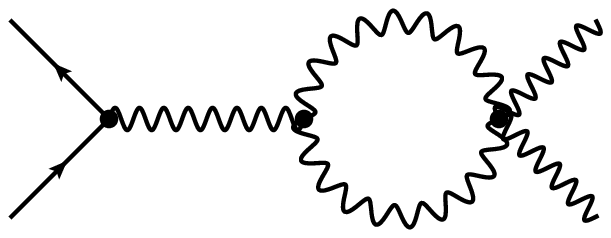} &
\includegraphics[height=1.5cm]{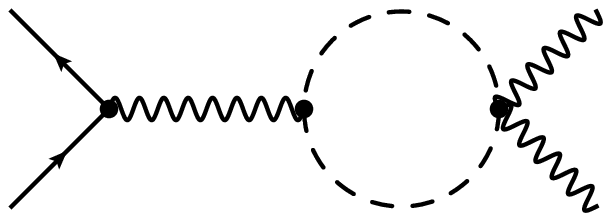} \\
$T2e$ & $T2f$ \\[10pt]
\includegraphics[height=1.5cm]{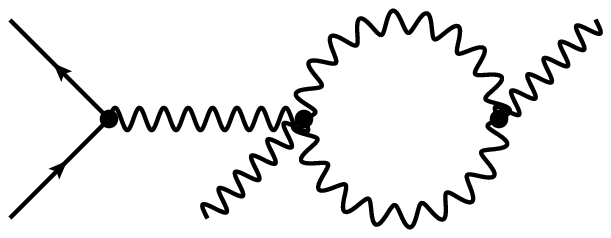} &
\includegraphics[height=1.5cm]{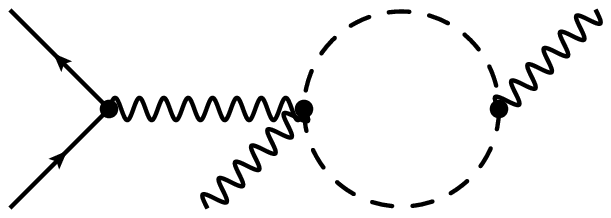} \\
$T2g$ & $T2h$ 
\end{tabular}
\end{center}
\caption{Diagrams of topology T2. See caption of Figure~\ref{fig:tree}.}\label{fig:t2}
\end{figure}

The diagrams of topology T2 are shown in Figure \ref{fig:t2}.

\subsubsection{T2a}

The sum of the ghost graph and its crossed-graph is
\begin{eqnarray}
\mathcal{C}(R_{T2a}) &=& \frac{g_i^4}{16\pi^2} \delta_{ij} i f_{dcf}^{(i)} f_{fbe}^{(i)} f_{ead}^{(i)} T^c_i\nn
\end{eqnarray}
\begin{eqnarray}
R_{T2a} &=& \frac{\mathcal{M}_1}{s}\Biggl[-\frac{1}{6\epsilon}+\frac{\Ls}{6}-\frac{11}{18}\Biggr]
\end{eqnarray}

\subsubsection{T2b}

The sum of the scalar graph and its crossed graph is
\begin{eqnarray}
\mathcal{C}(R_{T2b}) &=& \sum_k \frac{g_i^2 g_k^2}{16\pi^2} \delta_{ij} if_{abg}^{(i)}T^c_k \, \text{Tr}_{CS}\,\left(T^c_k T^g_i\right)  \nn
R_{T2b}  &=& \frac{\mathcal{M}_1}{s}\Biggl[\frac{2}{3\epsilon}-\frac{2\Ls}{3}+\frac{22}{9}\Biggr]\fs
\end{eqnarray}

If the gauge generators are orthogonal, then the $\text{Tr}_{CS}$ factor is proportional to $\delta_{ik}$. However, in general, the generators for $U(1)$ factors need not be orthogonal.

\subsubsection{T2c}

There is no crossed graph since the gauge bosons are real fields.
\begin{eqnarray}
\mathcal{C}(R_{T2c}) &=&  i\frac{g_i^4}{16\pi^2}\delta_{ij}  f_{cdf}^{(i)} f_{dae}^{(i)} f_{ebf}^{(i)} T^c_i\co\nn
\nn
R_{T2c}  &=& \frac{\mathcal{M}_1}{s}\Biggl[\frac{3}{\epsilon^2}+\frac{17-6\Ls}{2\epsilon}+ 
\frac32\Ls^2-\frac{17}{2}\Ls-\frac{\pi^2}{4}+\frac{95}{6}\Biggr]\fs\nn
\end{eqnarray}

\subsubsection{T2d}\label{sec:T2d}

The sum of graph T2d and its crossed graph is
\begin{eqnarray}
\mathcal{C}(R_{T2d}) &=& \sum_k \frac{g_i^2 g_k^2}{16\pi^2}i \delta_{ij} f^{(i)}_{abg} T^c_k \text{Tr}_{WF}\,\left(T^c_k T_i^g \right) \nn
R_{T2d}  &=& \frac{\mathcal{M}_1}{s}\Biggl[\frac{4}{3\epsilon}-\frac43\Ls+\frac{14}{9} \Biggr]
\end{eqnarray}

Graph T2d also has a piece proportional to the $\epsilon$ symbol, with a group theory factor proportional to 
$\text{Tr}_{WF}\,(T^c_k \left\{ T^b_j, T^a_i\right]\}$. This contribution is proportional to the gauge anomaly, must vanish when summed over all fermions in the loop for a consistent gauge theory, and so has not been given explicitly.
Our result for T2d differs from that in Ref.~\cite{bardin}. The formul\ae\ in Sec.~(14.13) give $26/9$ instead of $14/9$ for the finite part.

\subsubsection{T2e, T2f, T2g and T2h}

All these diagrams vanish.

\subsection{Topologies T3 and T4}

\begin{figure}
\begin{center}
\begin{tabular}{cc}
\includegraphics[height=1.5cm]{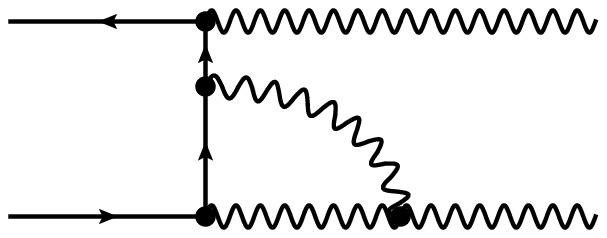}&\includegraphics[height=1.5cm]{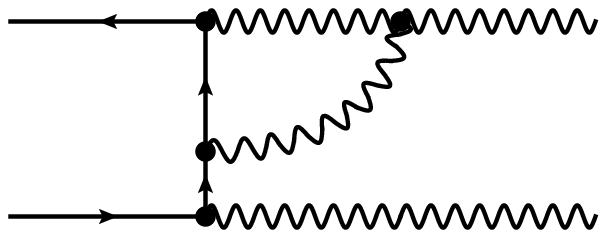}
\\
$T3a$ & $T4a$ \\[10pt]
\includegraphics[height=1.5cm]{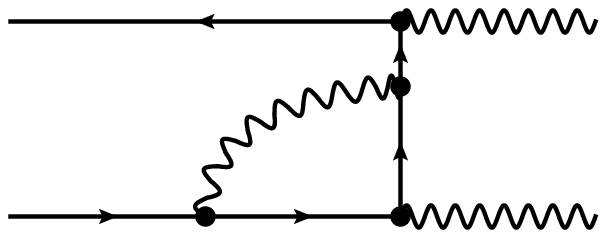} &
\includegraphics[height=1.5cm]{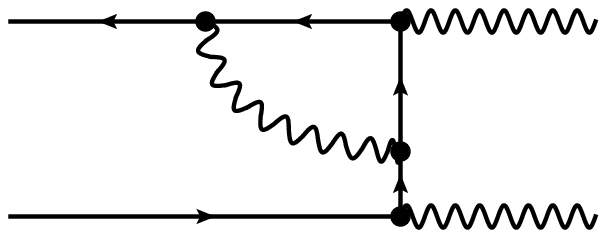} \\
$T3b$ & $T4b$ 
\end{tabular}
\end{center}
\caption{Diagrams of topology T3. See caption of Figure~\ref{fig:tree}.}\label{fig:t3}
\end{figure}

The diagrams with topology T3 and T4 are shown in Figure \ref{fig:t3}.

\subsubsection{T3a}
\begin{eqnarray}
\mathcal{C}(R_{T3a}) &=& \frac{g_i^3 g_j}{16\pi^2} \frac12 C_A(i) T^b_j T^a_i
\end{eqnarray}
\begin{eqnarray}
R_{T3a} &=& \frac{\mathcal{M}_0+2\mathcal{M}_1}{t}\left[\frac{2}{\epsilon^2}-\frac{2\Lt}{\epsilon}+\Lt^2-\frac{\pi^2}{6}\right]\nn
&+& \frac{\mathcal{M}_5}{t}\left[-\frac{2}{\epsilon^2}+\frac{2\Lt}{\epsilon}-\Lt^2+\frac{\pi^2}{6}+2\right]
\end{eqnarray}

The crossed graph $\bar R_{T3a}$ is given by applying Eq.~(\ref{mcross}) to $R_{3a}$, and has color factor
\begin{equation}
\mathcal{C}(\bar R_{T3a}) = \frac{g_i g_j^3}{16\pi^2} \frac12 C_A(j)  T^a_i T^b_j
\end{equation}
given from $\mathcal{C}(R_{T3a})$ by $i,a \leftrightarrow j,b$.

\subsubsection{T3b}

\begin{eqnarray}
\mathcal{C}(R_{T3b}) &=& \sum_k \frac{g_i g_j g_k^2}{16\pi^2} T^b_j T^c_k T^a_i T^c_k
\end{eqnarray}
\begin{eqnarray}
R_{T3b}&=& \frac{\mathcal{M}_0+2\mathcal{M}_1}{t}\left[\frac{1}{\epsilon}-\Lt+4\right]\nn
&+& \frac{\mathcal{M}_5}{t}\left[-\frac{4}{\epsilon}+4\Lt-10\right]
\end{eqnarray}

The crossed graph $\bar R_{T3b}$ is given by applying Eq.~(\ref{mcross}) to $R_{3b}$, and has color factor
\begin{equation}
\mathcal{C}(\bar R_{T3b}) = \sum_k \frac{g_i g_j g_k^2}{16\pi^2} T^i_a T^c_k T^b_j T^c_k 
\end{equation}
given from $\mathcal{C}(R_{T3b})$ by $i,a \leftrightarrow j,b$.

\subsubsection{T4a}
\begin{eqnarray}
\mathcal{C}(R_{T4a}) &=& \frac{g_i g_j^3}{16\pi^2} \frac12 C_A(j) T^b_j T^a_i
\end{eqnarray}
\begin{eqnarray}
R_{T4a} &=& \frac{\mathcal{M}_0+2\mathcal{M}_1}{t}\left[\frac{2}{\epsilon^2}-\frac{2\Lt}{\epsilon}+\Lt^2-\frac{\pi^2}{6}\right]\nn
&+& \frac{\mathcal{M}_4}{t}\left[-\frac{2}{\epsilon^2}+\frac{2\Lt}{\epsilon}-\Lt^2+\frac{\pi^2}{6}+2\right]
\end{eqnarray}

The crossed graph $\bar R_{T4a}$ is given by applying Eq.~(\ref{mcross}) to $R_{T4a}$, and has color factor
\begin{equation}
\mathcal{C}(\bar  R_{T4a}) = \frac{g_i^3 g_j}{16\pi^2} \frac12 C_A(i)  T^a_i T^b_j
\end{equation}
given from $\mathcal{C}(R_{T4a})$ by $i,a \leftrightarrow j,b$.

\subsubsection{T4b}

\begin{eqnarray}
\mathcal{C}(R_{T4b}) &=& \sum_k \frac{g_i g_j g_k^2}{16\pi^2} T^c_k T^b_j T^c_k T^a_i
\end{eqnarray}
\begin{eqnarray}
R_{T4b}&=& \frac{\mathcal{M}_0+2\mathcal{M}_1}{t}\left[\frac{1}{\epsilon}-\Lt+4\right]\nn
&+&\frac{\mathcal{M}_4}{t}\left[-\frac{4}{\epsilon}+4\Lt-10\right]
\end{eqnarray}

The crossed graph $\bar R_{T4b}$ is given by applying Eq.~(\ref{mcross}) to $R_{4b}$, and has color factor
\begin{equation}
\mathcal{C}(\bar R_{T4b}) = \sum_k \frac{g_i g_j g_k^2}{16\pi^2} T^c_k T^i_a T^c_k T^b_j
\end{equation}
given from $\mathcal{C}(R_{T4b})$ by $i,a \leftrightarrow j,b$.

\subsection{Topology T5}

\begin{figure}
\begin{center}
\begin{tabular}{cc}
\includegraphics[height=1.5cm]{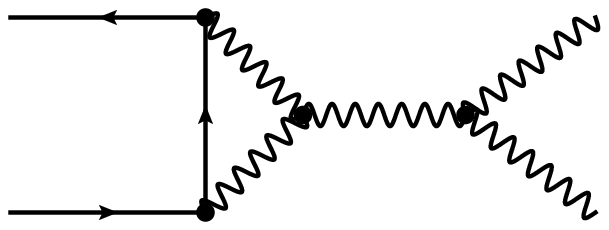}&\includegraphics[height=1.5cm]{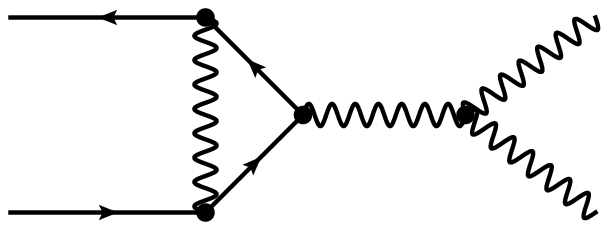}
\\
$T5a$ & $T5b$ 
\end{tabular}
\end{center}
\caption{Diagrams of topology T5. See caption of Figure~\ref{fig:tree}.}\label{fig:t7}
\end{figure}

The diagrams with topology T5 are shown in Figure~\ref{fig:t7}. There are no crossed graphs for this topology.

\subsubsection{T5a}

\begin{eqnarray}
\mathcal{C}(R_{T5a}) &=&  i f_{abc}^{(i)} \delta_{ij}\frac{g_i^4}{16\pi^2} \frac{C_A(i)}{2}  T_i^c \nn
R_{T5a}&=& \frac{\mathcal{M}_1}{s}   \left[- \frac{2}{\e}+ 2\LL_{s} -4  \right] 
\end{eqnarray}

\subsubsection{T5b}

\begin{eqnarray}
\mathcal{C}(R_{T5b}) &=&  \sum_{k} i f_{abc}^{(i)} \delta_{ij}\frac{g_i^2 g_k^2}{16\pi^2} T_k^d T_i^c
T_k^d \nn
R_{T5b} &=& \frac{\mathcal{M}_1}{s}\biggl[
  -\frac{4}{\e^2}-\frac{6}{\e}+\frac{4}{\e}\LL_{s}-2\LL_{s}^2 + 6\LL_{s}-16+\frac{\pi^2}{3}
  \biggr]\nn
\end{eqnarray}

\subsection{Topology T6}

\begin{figure}
\begin{center}
\begin{tabular}{cc}
\includegraphics[height=1.5cm]{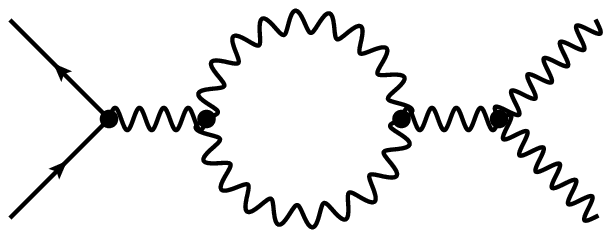}&\includegraphics[height=1.5cm]{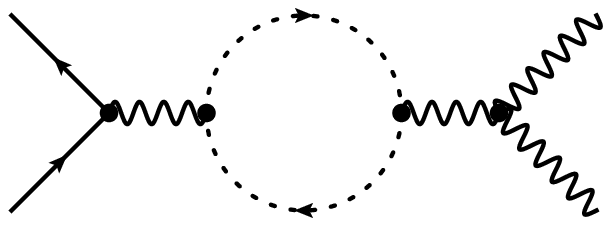}
\\
$T6a$ & $T6b$\\[10pt]
\includegraphics[height=1.5cm]{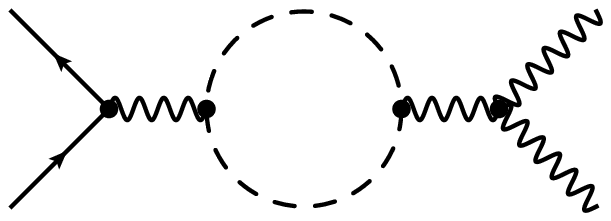}&\includegraphics[height=1.5cm]{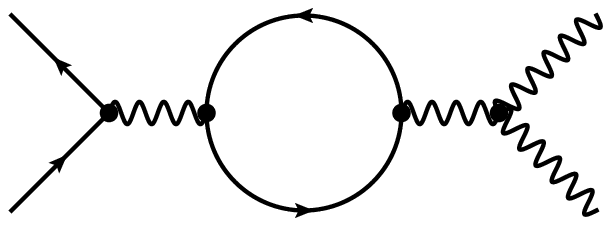}\\
$T6c$ & $T6d$  \\[10pt]
\includegraphics[height=1.5cm]{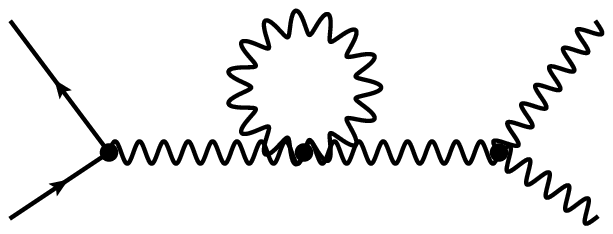}&\includegraphics[height=1.5cm]{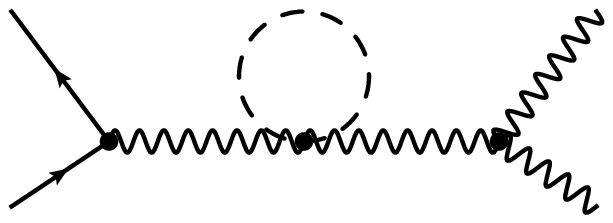}
\\
$T6e$ & $T6f$\\
\end{tabular}
\end{center}
\caption{Diagrams of topology T6. See caption of Figure~\ref{fig:tree}.}\label{fig:t8}
\end{figure}

The diagrams with topology T6 are shown in Figure~\ref{fig:t8}. There
are no crossed diagrams with this topology.

\subsubsection{T6a}

\begin{eqnarray}
\mathcal{C}(R_{T6a}) &=& \frac{g_i^4}{16\pi^2} if_{abc}^{(i)}\delta_{ij} C_A(i) T^c_i\nn
R_{T6a} &=& \frac{\mathcal{M}_1}{s} \biggl[  \frac{19}{6\epsilon} - \frac{19}{6} \Ls + \frac{58}{9}  \biggr] 
\end{eqnarray}

\subsubsection{T6b}

\begin{eqnarray}
\mathcal{C}(R_{T6b})&=& \frac{g_i^4}{16\pi^2} if_{abc}^{(i)}\delta_{ij} C_A(i) T^c_i\nn
R_{T6b} &=& \frac{\mathcal{M}_1}{s} \biggl[  \frac{1}{6\epsilon}  - \frac{1}{6} \Ls+ \frac{4}{9}  \biggr] 
\end{eqnarray}

\subsubsection{T6c}

\begin{eqnarray}
\mathcal{C}(R_{T6c})&=& \sum_k \frac{g_i^2 g_k^2}{16\pi^2}i f_{abc}^{(i)}\delta_{ij} \text{Tr}_{CS}(T^c_i T^d_k)T^d_k \nn
R_{T6c} &=& \frac{\mathcal{M}_1}{s}  \biggl[ -\frac{2}{3\epsilon}  +
  \frac{2}{3} \Ls - \frac{16}{9} \biggr] 
\end{eqnarray}

\subsubsection{T6d}

\begin{eqnarray}
\mathcal{C}(R_{T6d}) &=& \sum_k \frac{g_i^2 g_k^2}{16\pi^2}i f_{abc}^{(i)}\delta_{ij}  \text{Tr}_{WF}(T^c_i T^d_k)T^d_k \nn
R_{T6d} &=& \frac{\mathcal{M}_1}{s} \biggl[ -\frac{4}{3\epsilon}  +
  \frac{4}{3} \Ls- \frac{20}{9} \biggr]
\end{eqnarray}

\subsubsection{T6e, T6f}

These diagrams vanish in dimensional regularization.

\subsection{Topology T7}

\begin{figure}
\begin{center}
\begin{tabular}{cc}
\includegraphics[height=1.5cm]{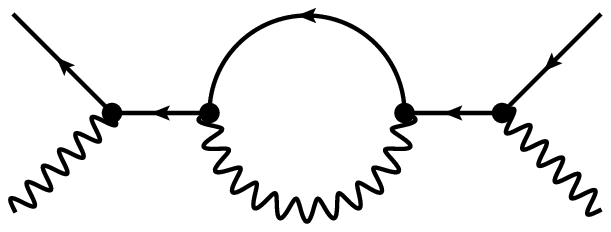}& \\
$T7$ & 
\end{tabular}
\end{center}
\caption{Diagram with topology T7. See caption of Figure~\ref{fig:tree}.}\label{fig:t9}
\end{figure}

The diagram with topology T7 is shown in Figure~\ref{fig:t9}.

\begin{eqnarray}
\mathcal{C}(R_{T7}) &=&  \sum_k \frac{g_i g_jg_k^2}{16\pi^2} T^b_j T_k^c T_k^c
T^a_i \nn
R_{T7} &=& \frac{\mathcal{M}_0 +2
\mathcal{M}_1}{t}\left[\frac{1}{\epsilon} - \Lt +1 \right] 
\end{eqnarray}

The crossed graph $\bar R_{T7}$ is given by applying Eq.~(\ref{mcross}) to $R_{T7}$, and has color factor
\begin{equation}
\mathcal{C}(\bar R_{T7}) = \sum_k \frac{g_i g_jg_k^2}{16\pi^2} T^a_i T_k^c T_k^c T^b_j 
\end{equation}
given from $\mathcal{C}(R_{T7})$ by $i,a \leftrightarrow j,b$.

\section{$\bar{q} q \to VV$ in the Standard Model}\label{sec:smvv}

The group theory factors for $\bar{q} q \to V^a_i V^b_j$ have been written in a form where they are applicable to gauge boson production for an arbitrary product group, with fermions in an arbitrary irreducible representation. In this section, we tabulate the group theory factors and give their values for the standard model.

The group theory factors when the two vector bosons belong to the same group $G_i$ are given in Table~\ref{tab:VV}. 
The only assumption we have made on the structure of the gauge theory is that the gauge generators are orthogonal,
\begin{eqnarray}
\text{Tr}_{CS} \left(T^a_i T^b_j\right) &=& \delta_{ab} \delta_{ij} \trcs{i}\co\nn
\text{Tr}_{WF} \left(T^a_i T^b_j\right) &=& \delta_{ab} \delta_{ij} \trwf{i}\co
\label{eq57}
\end{eqnarray}
which define $\trcs{i}$ and $\trwf{i}$. The orthogonality only needs to be checked if both $i$ and $j$ correspond to $U(1)$ factors, and is satisfied in theories which arise as the low-energy limit of unified theories based on semisimple Lie groups. The group factors and coupling constants have to be evaluated using their values for $G_i$ in the representation of the fermion.

The group theory factors have been written in terms of
\begin{eqnarray}
\ga &=& i f_{abc}T^c\co \nn 
\GAC&=& \frac12 \left\{T^a,T^b\right\}\co \nn
\GFF &=& \frac12 f_{acg}f_{bch}\left\{T^g, T^h\right\}\fs
\end{eqnarray}
$\GAC$ and $\GFF$ can not, in general, be written as expressions linear in the group generators $T^a$.
For $SU(N)$ groups  in the fundamental representation,
\begin{eqnarray}
\GAC&=& \frac12 d_{abc} T^c + \frac{1}{2N}\delta_{ab} \co \nn
\GFF &=& \frac14 C_A d_{abc} T^c + \frac1 2 \delta_{ab}\co
\label{sunform}
\end{eqnarray}
but these expressions are not valid for general $SU(N)$ representations. Some examples of $\GFF$ for higher $SU(N)$ representations are given in Appendix~A of Ref.~\cite{Dashen:1994qi}.
For  $U(1)$ groups, 
\begin{eqnarray}
\ga &=& 0\co\nn
\GAC&=& Y_i^2 \co \nn
\GFF &=& 0 \co
\label{u1form}
\end{eqnarray}
where $Y_i$ is the $U(1)$ charge. 

$\X$ is defined by
\begin{eqnarray}
\X &=& \sum_i \alpha_i C_{F,Q}(i)
\label{eq58}
\end{eqnarray}
where $C_{F,Q}(i)$ is the quadratic Casimir of the incoming fermion under gauge group $G_i$, and the sum is over all gauge groups. $C_A$ is the Casimir in the adjoint representation.

For the standard model, the high-energy amplitude is most conveniently written in terms of the gauge bosons of the unbroken gauge theory --- $W^a$ of $SU(2)$, $B$ of $U(1)$ and gluons $G^a$ of $SU(3)$, and we can use Eqs.~(\ref{sunform},\ref{u1form}) with $Y_i \to Y_Q=1/6$ and $N=2,3$. The $d$-symbol vanishes for $SU(2)$. The factors in Eqs.~(\ref{eq57},\ref{eq58}) are
\begin{eqnarray}
\trcs{i}&=& \left\{ \begin{array}{cc} 
0 & \text{for}\quad SU(3) \\
\frac12 n_S& \text{for}\quad SU(2) \\
\frac12n_S & \text{for}\quad  U(1) \co
  \end{array} \right.  \nn 
 \trwf{i} &=& \left\{ \begin{array}{cc} 
2n_g & \text{for}\quad SU(3) \\
2n_g& \text{for}\quad SU(2) \\
\frac{10}{3}n_g& \text{for}\quad  U(1) \co
  \end{array} \right. 
\end{eqnarray}
and
\begin{eqnarray}\label{eq:lambda}
\X &=& \frac43 \alpha_3 + \frac34 \alpha_2 +Y_Q^2 \alpha_1
\end{eqnarray}
where $n_g=3$ is the number of fermion generations, and $n_S=1$ is the number of Higgs doublets.

Group theory factors for the crossed graphs are given by $a \leftrightarrow b$, so that $\ga \to -\ga$ changes sign, $\GAC \to \GAC$ and $\GFF \to \GFF$.

\begin{table*}
\begin{eqnarray*}
\renewcommand{\arraystretch}{1.8} 
\begin{array}{cc|cc|cc|cc}
\hline
R_1 & 4\pi \alpha  \left(\GAC- \frac12\ga\right)  & 
R_2 & -4\pi \alpha  \ga & 
&&
& \\

T1a &  -\alpha^2\left(\GFF-\frac14 C_A \ga \right)&
T1b & -\frac12 \alpha^2 C_A \GAC + \alpha^2 \GFF&
T1c & \alpha\left(\frac14 \alpha C_A -\frac12\X\right)\ga +\alpha^2 \GFF &
T1d & 2\alpha^2 \GFF \\[-5pt]
&&
&&
& +\alpha \left( \X -\alpha C_A\right)\GAC &
& \\

T2a &  \frac12\alpha^2 C_A\ga &
T2b & \alpha^2  \trcs i \ga   &
T2c & - \frac12 \alpha^2 C_A\ga &
T2d & \alpha^2 \trwf i \ga \\

T3a &  \frac12 \alpha^2 C_A \left(\GAC- \frac12\ga\right)  &
T3b & \alpha \left( \GAC- \frac12\ga\right) \left[\X-\frac12 \alpha C_A   \right] &
T4a & \frac12 \alpha^2 C_A\left( \GAC- \frac12\ga\right)  &
T4b & \alpha\left(\GAC- \frac12\ga\right)  \left[\X - \frac12 \alpha  C_A \right]    \\

T5a & \frac12 \alpha^2 C_A \ga &
T5b & \alpha \ga \left[ \X - \frac12 \alpha  C_A \right]  &  & & &  \\

T6a & \alpha^2 C_A\ga  & 
T6b &  \alpha^2 C_A\ga  &
T6c & \alpha^2 \trcs i  \ga  &
T6d &  \alpha^2  \trwf  i \ga \\

T7 & \alpha \left(\GAC- \frac12\ga\right)  \X  & & & & & & \\
\hline
\end{array}
\end{eqnarray*}
\caption{Group theory coefficients $\mathcal{C}_i$ for the production of two
  identical gauge bosons.  The
  coefficients $\bar{\mathcal{C}}_i$ of the crossed diagrams are given by
  $\mathcal{C}_i$ with  $a \leftrightarrow
  b$, under which $\ga \to -\ga$, $\GAC \to \GAC$ and $\GFF \to \GFF$. The notation is explained in the main text.}\label{tab:VV}
\end{table*}

Group theory factors for the production of gauge bosons in two different gauge groups are given in Table~\ref{tab:GW}.
The gauge bosons with momentum $p_4$ and $p_3$ are $V^a_i$ and $V^b_j$, respectively. We define
\begin{eqnarray}
\al = \sqrt{\alpha_{i} \alpha_{j}}\ T^a_i T^b_j\fs
\end{eqnarray}
The factors for the crossed graphs are given by $i \leftrightarrow j$. This gives the group theory factors for the reactions $\bar{q} q \to G^a(p_4) W^b(p_3)$, $\bar{q}
q \to G^a(p_4) B(p_3)$ and $\bar{q} q \to W^a(p_4) B(p_3)$. The reaction $\bar{q} q \to W^a(p_3) G^b(p_4)$,
is related to $\bar{q} q \to G^a(p_4) W^b(p_3)$ by exchanging the final gauge bosons, i.e.\ by the swap
$i,a,p_4 \leftrightarrow j,b,p_3$.

\begin{table*}
\begin{eqnarray*}
\renewcommand{\arraystretch}{1.8} 
\begin{array}{cc|cc|cc|cc}
\hline
R_1 & 4\pi \al  & 
R_2 & 0 & 
&&
& \\

T1a &  0 &
T1b & -\frac12\al \alpha_{i}  C_A(i) &
T1c & \al \left[ \X-  \frac12\alpha_{i}
 C_A(i)-\frac12 \alpha_{j}  C_A(j) \right]  &
T1d & 0 \\

T2a & 0  &
T2b & 0 &
T2c &  0  &
T2d & 0 \\

T3a &\frac12  \al  \alpha_{i} C_A(i) &
T3b & \al  \left[
  \X- \frac12 \alpha_{i} C_A(i) \right] &
T4a &\frac12   \al \alpha_{j} C_A(j) &
T4b &  \al \left[
  \X- \frac12 \alpha_{j} C_A(j) \right]  \\

T5a & 0 & T5b & 0 & & & & \\

T6a & 0 &
T6b & 0 &
T6c & 0 &
T6d & 0 \\

T7 & \al \X & & & & & &\\

\hline
\end{array}
\end{eqnarray*}
\caption{Group theory coefficients $\mathcal{C}_i$ for the production of two
  different gauge bosons. Here $\al = \sqrt{\alpha_{i} \alpha_{j}} T^a_i T^b_j$. The
  coefficients $\bar{\mathcal{C}}_i$ of the crossed diagrams are given by
  $\mathcal{C}_i$ with  $i \leftrightarrow j$.}\label{tab:GW}
\end{table*}

\section{Scalar production}\label{sec:scalar}

The notation for the scalar production is analogous to that for vector boson production. The full
amplitude is given by the sum of all diagrams $S_i$ with 
group theory factor $\mathcal{C}(S_i)$.
\begin{equation}\label{eq:amps}
\mathcal{M}  =  \sum_i \mathcal{C}(S_i)S_i\fs
\end{equation}
As in the vector boson case, the $\bar S$ and $\bar \mathcal{C}$ denote the crossed
diagrams and group theory factors. 
The Dirac matrix element is 
\begin{eqnarray}
\mphi = \bar{v}(p_2)\slashed{p}_4 P_L u(p_1)\fs
\end{eqnarray}

The diagrams are classified in terms of the topologies
given in Figure~\ref{fig:topos}.  Exchanging the two final state scalars gives
\begin{eqnarray}
\mphi &\leftrightarrow& -\mphi\co\nn
t &\leftrightarrow& u\fs
\label{scross}
\end{eqnarray}

\subsection{Tree level amplitude}

\begin{figure}
\begin{center}
\begin{tabular}{cc}
\includegraphics[height=1.5cm]{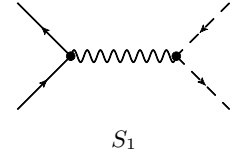}& \\
$S_{1}$ & 
\end{tabular}
\end{center}
\caption{Tree level diagram for the production of two scalars. See caption of Figure~\ref{fig:tree}.}\label{fig:trees}
\end{figure}

The tree level diagram is shown in Figure~\ref{fig:trees}.

At tree level, one finds 
\begin{eqnarray}
\mathcal{C}(S_1) &=& \sum_i g_i^2 T^a_i \otimes T^a_i \co \nn
S_1 &=&  \frac{2}{s} \, \mphi \fs
\end{eqnarray}

For the group theory factor $X \otimes Y$, $X$ acts on the initial fermion space and $Y$ on the final scalar particle space.

\subsection{Topology T1}

\begin{figure}
\begin{center}
\begin{tabular}{cc}
\includegraphics[height=1.5cm]{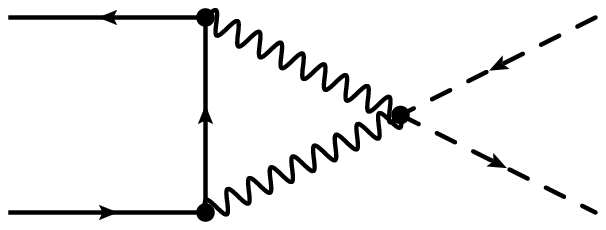}&\includegraphics[height=1.5cm]{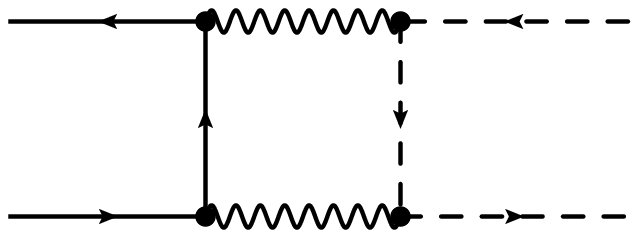} \\
$T1a$ & $T1b$
\end{tabular}
\end{center}
\caption{Diagram with topology T1. See caption of Figure~\ref{fig:tree}.}\label{fig:t1s}
\end{figure}

The diagrams of topology T1 are shown in Figure~\ref{fig:t1s}.

\subsubsection{T1a}

This diagram vanishes.

\subsubsection{T1b}

\begin{eqnarray}
\mathcal{C}(S_{T1b}) &=& \sum_{ij}  \frac{g_i^2 g_j^2}{16\pi^2} T^a_i T^b_j \otimes
T^b_j T^a_i \nn
S_{T1b} &=& \mphi  \Biggl[ -\frac{9}{s \e^2}+\frac{1}{s
    \e}\left(\Ls+8\Lt-2\right)
  -\Ls^2\frac{1}{2u}\left(\frac{7t}{s}+3\right)\nn
&&+\frac{2}{u}\Lt^2+\Ls\Lt\frac{4}{u}\left(\frac{t-u}{s}\right) +\Ls\frac{2}{s}-\frac{4}{s}
\nn &&-\frac{\pi^2}{4u}\left(11+\frac{19t}{s}\right)\Biggr] 
\end{eqnarray}

The box diagram $S_{T1b}$ is the only one where a crossed diagram
exists for scalar production. Exchanging the final state scalars gives the crossed-box
graph. The amplitude is given by applying Eq.~(\ref{scross}) to $S_{T1b}$.
For the group theory factor, one finds
\begin{eqnarray}
\mathcal{C}(\bar S_{T1b}) &=& \sum_{ij}  \frac{g_i^2 g_j^2}{16\pi^2} T^a_i T^b_j
\otimes T^a_i T^b_j  \fs
\end{eqnarray}

\subsection{Topology T2}

\begin{figure}
\begin{center}
\begin{tabular}{cc}
\includegraphics[height=1.5cm]{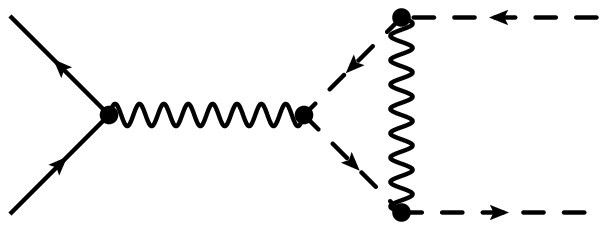}&\includegraphics[height=1.5cm]{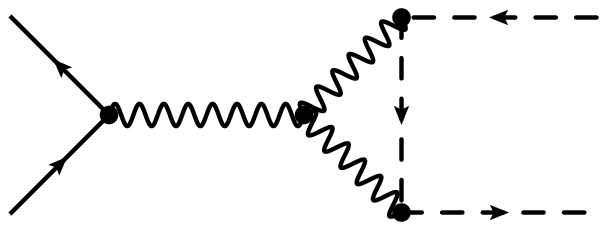} \\
$T2a$ & $T2b$ \\[10pt]
\includegraphics[height=1.5cm]{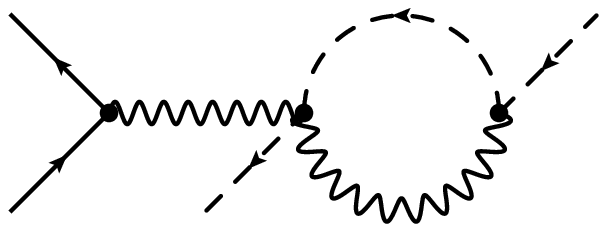}&\includegraphics[height=1.5cm]{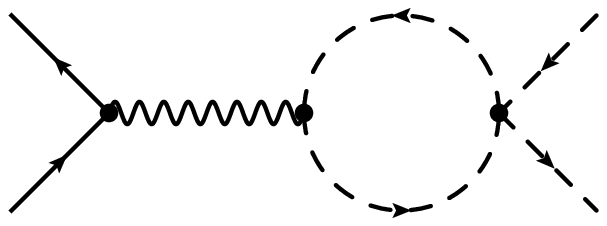} \\
$T2c$ & $T2d$ \\[10pt]
\includegraphics[height=1.5cm]{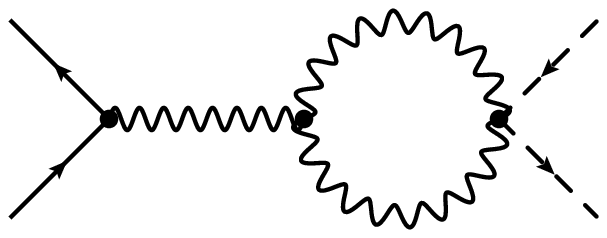} &  \\
$T2e$ & 
\end{tabular}
\end{center}
\caption{Diagram with topology T2. See caption of Figure~\ref{fig:tree}.}\label{fig:t2s}
\end{figure}

The diagrams of topology T2 are shown in Figure~\ref{fig:t2s}. There are no crossed diagrams with this topology.

\subsubsection{T2a}

\begin{eqnarray}
\mathcal{C}(S_{T2a}) &=& \sum_{i,j} \frac{g_i^2 g_j^2}{16\pi^2} T^a_i \otimes T_j^b
T^a_i T_j^b\\
S_{T2a} &=&  \frac{ \mphi}{s}
\biggl[-\frac{4}{\e^2}-\frac{8}{\e}+\frac{4}{\e}\LL_{s}-2\LL_{s}^2  + 8\LL_{s}-16+\frac{\pi^2}{3}
\biggr] \nonumber
\end{eqnarray}

\subsubsection{T2b}

\begin{eqnarray}
\mathcal{C}(S_{T2b}) &=&  \sum_{i} \frac{g_i^4 }{16\pi^2} \frac{C_A(i)}{2} T^a_i
\otimes  T^a_i \\
S_{T2b} &=& \frac{\mphi}{s} \biggl[\frac{1}{\e^2}-\frac{2}{\e}
  -\frac{1}{\e}\LL_{s}+\frac12\LL_{s}^2  + 2\LL_{s}-4-\frac{\pi^2}{12} \biggr] \nonumber
\end{eqnarray}

\subsubsection{T2c, T2d, T2e}

These diagrams vanish. Graph T2d is the only diagram involving the $\lambda \phi^4$ coupling.

\subsection{Topologies T3, T4}

There are no diagrams with topology T3 and T4.

\subsection{Topology T5}

\begin{figure}
\begin{center}
\begin{tabular}{cc}
\includegraphics[height=1.5cm]{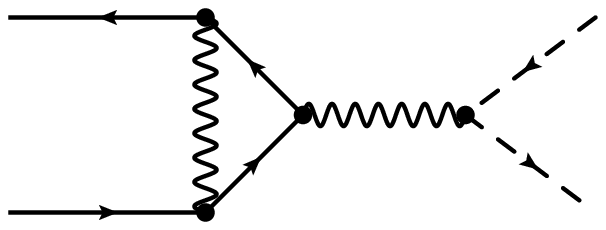}&\includegraphics[height=1.5cm]{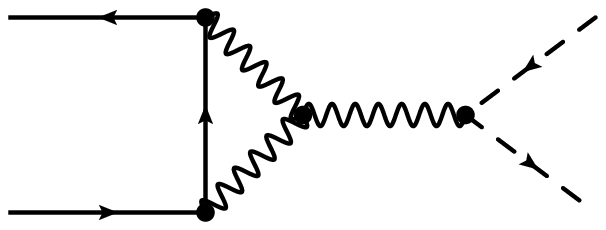} \\
$T5a$ & $T5b$
\end{tabular}
\end{center}
\caption{Diagram with topology T5. See caption of
  Figure~\ref{fig:tree}.}\label{fig:t5s}
\end{figure}

The diagrams of topology T5 are shown in Figure~\ref{fig:t5s}. There are no crossed graphs with this topology.

\subsubsection{T5a}

\begin{eqnarray}
\mathcal{C}(S_{T5a}) &=& \sum_{i,j} \frac{g_i^2 g_j^2}{16\pi^2} T^b_i T^a_j T^b_i \otimes T^a_j \co \\
S_{T5a} &=& \frac{\mphi}{s}\left[-\frac{4}{\epsilon ^2} +\frac{-6+4
    \Ls}{ \epsilon }-2 \Ls^2+6 \Ls+\frac{\pi ^2}{3 }-16 \right] \nonumber
\end{eqnarray}

\subsubsection{T5b}

\begin{eqnarray}
\mathcal{C}(S_{T5b}) &=& \sum_k \frac{g_i^4}{16\pi^2}\frac 12 C_A(i) T^a_i  \otimes T^c_i \nn
S_{T5b} &=& \frac{\mphi}{s}\left[-\frac{2}{ \epsilon }+2\Ls -4 \right] 
\end{eqnarray}

\subsection{Topology T6}

\begin{figure}
\begin{center}
\begin{tabular}{cc}
\includegraphics[height=1.5cm]{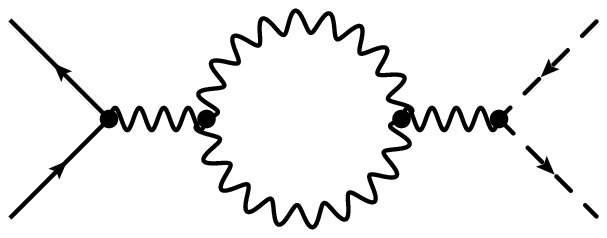}&\includegraphics[height=1.5cm]{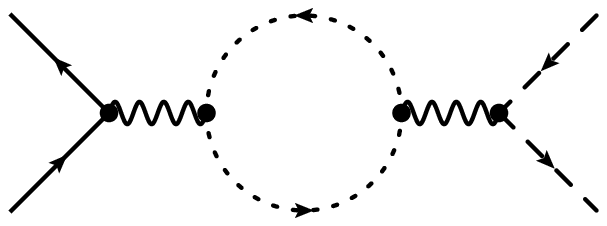} \\
$T6a$ & $T6b$\\[10pt]
\includegraphics[height=1.5cm]{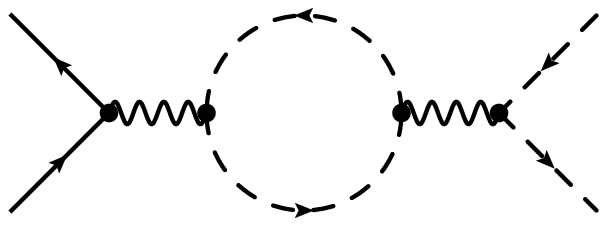}&\includegraphics[height=1.5cm]{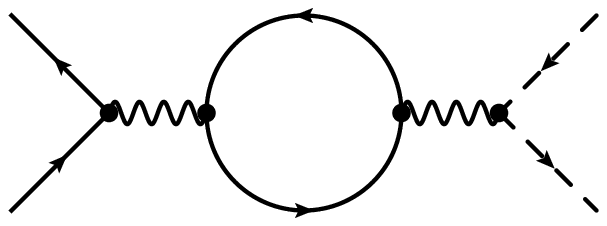} \\
$T6c$ & $T6d$
\end{tabular}
\end{center}
\caption{Diagram with topology T5. See caption of
  Figure~\ref{fig:tree}.}\label{fig:t6s}
\end{figure}

The diagrams of topology T6 are shown in Figure~\ref{fig:t6s}. There are no crossed graphs with this topology.

\subsubsection{T6a}

\begin{eqnarray}
\mathcal{C}(S_{T6a}) &=& \frac{g_i^4}{16\pi^2} C_A(i) T_i^a \otimes T_i^a \nn
S_{T6a} &=& \frac{\mphi}{s} \biggl[  \frac{19}{6\epsilon} - \frac{19}{6} \Ls + \frac{58}{9}  \biggr] 
\end{eqnarray}

\subsubsection{T6b}

\begin{eqnarray}
\mathcal{C}(S_{T6b})  &=& \frac{g_i^4}{16\pi^2} C_A(i) T_i^a \otimes T_i^a \nn
S_{T6b} &=& \frac{\mphi}{s} \biggl[  \frac{1}{6\epsilon}- \frac{1}{6} \Ls  + \frac{4}{9}  \biggr] 
\end{eqnarray}

\subsubsection{T6c}

\begin{eqnarray}
\mathcal{C}(S_{T6c}) &=& \sum_k \frac{g_i^2 g_k^2}{16\pi^2} T_i^a \otimes T^d_k\text{Tr}_{CS}(T^a_i T^d_k) \nn
S_{T6c} &=& \frac{\mphi}{s}  \biggl[ -\frac{2}{3\epsilon}  + \frac{2}{3} \Ls- \frac{16}{9} \biggr] 
\end{eqnarray}

\subsubsection{T6d}

\begin{eqnarray}
\mathcal{C}(S_{T6d})  &=& \sum_k \frac{g_i^2 g_k^2}{16\pi^2} T_i^a \otimes T^d_k \text{Tr}_{WF}(T^a_i T^d_k)\nn
S_{T6d} &=& \frac{\mphi}{s} \biggl[ -\frac{4}{3\epsilon} - \frac{20}{9} +
  \frac{4}{3} \Ls \biggr]
\end{eqnarray}

\section{$q \bar{q}  \to \phi^\dagger \phi$ in the Standard Model}\label{sec:smscalar}

Group theory factors for scalar production are given in Table~\ref{tab:ss}. They depend on group invariants which are listed below, followed by their values in the standard model.
\begin{eqnarray}
\ssab &=& \sum_i \alpha_i \alpha_j T_i^a T_j^b  \otimes T_j^b T_i^a \nn
&=& \left[\frac12\alpha_2^2 +2 \alpha_1 \alpha_2 Y_Q Y_\phi\right] t^a \otimes t^a\nn
&&+ \left[ \frac{3}{16} \alpha_2^2 +\alpha_1^2 Y_Q^2 Y_\phi^2\right] \openone \otimes \openone  \nn
\ssba &=&   \sum_i \alpha_i \alpha_j T_i^a T_j^b  \otimes  T_i^a T_j^b \nn
&=& \left[-\frac12\alpha_2^2 +2 \alpha_1 \alpha_2 Y_Q Y_\phi\right] t^a \otimes t^a\nn
&& + \left[ \frac{3}{16} \alpha_2^2 +\alpha_1^2 Y_Q^2 Y_\phi^2\right] \openone \otimes \openone \nn
\ssi &=& \sum_i \alpha_i T_i^a \otimes T_i^a \nn
&=& \alpha_2 t^a \otimes t^a +\alpha_1  Y_Q Y_\phi \openone \otimes \openone \nn
\ssii &=&  \sum_i \alpha_i^2 C_A(i) T_i^a \otimes T_i^a = 2 \alpha_2^2 t^a \otimes t^a \nn
\sscs &=& \sum_i \alpha_i^2 \trcs{i} T_i^a \otimes T_i^a \nn
&=& \frac12\alpha_2^2 n_S t^a \otimes t^a + 2 \alpha_1^2  n_S Y_Q Y_\phi^3 \openone \otimes \openone \nn
\sswf &=& \sum_i \alpha_i^2 \trwf{i} T_i^a \otimes T_i^a \nn
&=& 2\alpha_2^2 n_g  t^a \otimes t^a + \frac{10}{3}\alpha_1^2 n_g Y_Q Y_\phi \openone \otimes \openone \nn
\lamQ &=& \sum_i \alpha_i  C_{F,Q}(i) = \frac43 \alpha_3+  \frac{3}{4} \alpha_2 + \alpha_1 Y_Q^2 \nn
\lamphi &=& \sum_i \alpha_i C_{F,\phi}(i) = \frac{3}{4} \alpha_2 + \alpha_1 Y_\phi^2
\end{eqnarray}
Here $Y_\phi=1/2$ is the hypercharge of the Higgs scalar, $C_{F,\phi}$ is the Casimir in the representation of the scalar field, and $\trcs{i}$ and $\trwf{i}$ are defined in Eq.~(\ref{eq57}).

\subsection{Top quark loops}

Top quark loops have to be included in the high scale matching for scalar production in the standard model, since
the top-quark Yukawa coupling is comparable to the gauge couplings.

Since this contribution to the high-scale matching depends on the
details of the theory, top quark loops are only computed for
the standard model. Here $y_t$ is the top quark
Yukawa coupling and $Y(t_L)$ and $Y(t_R)$ are used for the $U(1)$
charges of the left- and right-handed top quarks, respectively. Note
$Y(t_R)-Y(t_L)=Y_\phi=1/2$.

\begin{figure}
\begin{center}
\begin{tabular}{cc}
\includegraphics[height=1.5cm]{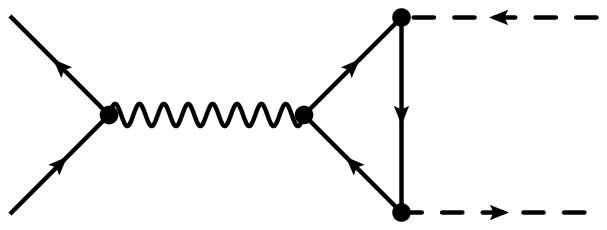}&  \\
$S_{\text{top}}$&
\end{tabular}
\end{center}
\caption{Top quark loop contribution to the amplitude. See also caption of 
  Figure~\ref{fig:tree}.}\label{fig:top}
\end{figure}

The relevant diagram is shown in Figure~\ref{fig:top}, and has to be added to the amplitude for
scalar production, Eq.~(\ref{eq:amps}), The graph with $t_L$ coupling to the gauge boson is
\begin{eqnarray}
&&3y_t^2  \frac{\alpha_1}{4\pi}Y_Q Y(t_L)\frac{\mphi}{s} \Biggl[-\frac{2}{\e} +2 \Ls -4\Biggr]\
\openone \otimes \openone \nn
&&-3y_t^2 \frac{\alpha_2}{4\pi} \frac{\mphi}{s}\Biggl[-\frac{2}{\e}+2 \Ls -4\Biggr]  
t^a \otimes t^a
\end{eqnarray}
and the graph with $t_R$ coupling to the gauge boson is
\begin{eqnarray}
- 3y_t^2\frac{\alpha_1}{4\pi } Y_Q Y(t_R) \frac{\mphi}{s}\Biggl[-\frac{2}{\e}+2 \Ls -4\Biggr]
\openone \otimes \openone 
\end{eqnarray}
\begin{eqnarray}
S_{\text{top}} &=& -3y_t^2\frac{ \alpha_1 }{4\pi} Y_Q Y_\phi
\frac{\mphi}{s}\Biggl[-\frac{2}{\epsilon} +2 \Ls -4\Biggr]\openone \otimes \openone \nn
&&-3y_t^2\frac{ \alpha_2 }{4\pi} \frac{\mphi}{s}\Biggl[-\frac{2}{\epsilon} +2
  \Ls -4\Biggr] t^a \otimes t^a \fs
  \label{tloop}
\end{eqnarray}
using $Y(t_R)-Y(t_L)=Y_\phi$.

\begin{table}
\begin{eqnarray*}
\renewcommand{\arraystretch}{1.8} 
\begin{array}{cc|cc}
\hline
S_1 & 4\pi \ssi & \\

T1b & \ssab &

\bar T1b & \ssba \\

T2a &  -\frac12 \ssii + \lamphi \ssi&
T2b & \frac12 \ssii \\ 

T5a & -\frac12 \ssii +\lamQ \ssi &
T5b &  \frac12 \ssii \\

T6a & \ssii &
T6b & \ssii \\
T6c &\sscs &
T6d & \sswf  \\

\hline
\end{array}
\end{eqnarray*}
\caption{Group theory coefficients for the production of two
  charged scalars. The notation is explained in the main text.}\label{tab:ss}
\end{table}

\section{Consistency Checks}\label{sec:check}

There is a consistency check on our matching coefficients, which follows from the fact that $S$-matrix elements
are independent of the scale $\mu$ at which one matches from the full theory to SCET. Consider, for example
electroweak gauge boson production by left-handed quark doublets. There are five SCET operators which contribute~\cite{Chiu:2009ft,Chiu:2009mg}
\begin{eqnarray}
O_1 &=& \bar Q^{(u)}_2 Q^{(u)}_1 W^a_4W^a_3\nn
O_2 &=& \bar Q^{(u)}_2 t^c Q^{(u)}_1  i \epsilon^{abc} W^a_4 W^b_3 \nn
O_3 &=& \bar Q^{(u)}_2 t^a Q^{(u)}_1  B_4 W^a_3\nn
O_4 &=& \bar Q^{(u)}_2 t^a Q^{(u)}_1 W^a_4 B_3 \nn
O_5 &=& \bar Q^{(u)}_2  Q^{(u)}_1 B_4 B_3 
\label{185.b}
\end{eqnarray}
with matching coefficients $C_i(\mu)$ at the matching scale $\mu$. Write the coefficients as
\begin{eqnarray}
C_i &=& C_i^{(0)} + C_i^{(1)} + \ldots
\end{eqnarray}
where $ C_i^{(0)}$ are the tree-level coefficients, $ C_i^{(1)}$ are the one-loop contributions, etc.
Then $\mu$-independence implies the constraint
\begin{eqnarray}
\sum_k \mu \frac{{\rm d}\alpha_k }{{\rm d}\mu} \frac{\partial C_i^{(0)}}{\partial \alpha_k} + \mu \frac{\partial C_i^{(1)}}{\partial \mu} &=& \gamma^{(1)}_{ij} C_j^{(0)}
\label{conscond}
\end{eqnarray}
where the sum on $k$ is over the three standard model gauge groups, and $\gamma^{(1)}_{ij} $ is the one-loop anomalous dimension in SCET computed in Refs.~\cite{Chiu:2009ft,Chiu:2009mg}.

The SCET anomalous dimension is
\begin{eqnarray}
\bm{\gamma} &=& \left(2 \gamma_Q+ 2 \gamma_V\right) +
\bm{\gamma}_S
\end{eqnarray}
where $\gamma_Q$ and $\gamma_V=\gamma_{W,B}$ are the collinear anomalous dimensions of $Q$, $W$, and $B$, and $\gamma_S$ is the soft anomalous dimension~\cite{Chiu:2009ft,Chiu:2009mg}. Using the values
in Refs.~\cite{Chiu:2009ft,Chiu:2009mg} gives
\begin{widetext}

The anomalous dimension is  ($\LL=\log s/\mu^2$)
\begin{eqnarray}
\bm{\gamma} &=& 2 \gamma_Q \openone + \left[ \begin{array}{ccccc}
2 \gamma_W & 0 & 0 & 0 & 0 \\
0 &  2 \gamma_W  & 0 &0 & 0\\
0 & 0 & \gamma_W+\gamma_B & 0 & 0 \\
0 & 0 & 0 & \gamma_W+\gamma_B & 0  \\
0 & 0 & 0 & 0 & 2 \gamma_B  \\
\end{array} \right]\nn
&&+ \frac{\alpha_1}{\pi} \left(-i\pi Y_Q^2\right)+\frac{\alpha_s}{\pi}\left(-\frac43i \pi\right) +\frac{\alpha_2}{\pi}\left[ \begin{array}{ccccc}
-\frac{11}{4}i \pi & U-T & 0 & 0 & 0 \\
2(U-T) & -\frac{11}{4}i \pi +T+U& 0 &0 & 0\\
0 & 0 & -\frac{7}{4}i \pi + T + U & 0 & 0 \\
0 & 0 & 0 & -\frac{7}{4}i \pi + T + U & 0  \\
0 & 0 & 0 & 0 & -\frac{3}{4}i \pi   \\
\end{array} \right]
\label{adW}
\end{eqnarray}
 where the first line is the collinear contribution, and the second line is the soft contribution. 
 
 \end{widetext}
 
Here
\begin{eqnarray}
\gamma_Q &=& \left( \frac{\alpha_s}{4\pi} \frac43 +\frac{\alpha_2}{4\pi} \frac34  +\frac{\alpha_1}{4\pi} Y_Q^2  \right) \left( 2 \log \frac{s}{\mu^2}-3\right)\nn
\gamma_W &=& \frac{\alpha_2}{4\pi}  \left( 4 \log \frac{s}{\mu^2}-\frac{19}{6}\right)\nn
\gamma_B &=& \frac{\alpha_1}{4\pi}  \left( \frac{41}{6}\right)\nn
\end{eqnarray}
$T=\log(-t/s)-i \pi$, $U=\log(-u/s)-i\pi$, and $Y_Q=1/6$.

The consistency condition Eq.~(\ref{conscond}) is satisfied using our results for the matching coefficients and Eq.~(\ref{adW}). This only checks the relation between the $\log \mu$ terms at one-loop and the tree-level coefficients.

For scalar production by doublet quarks, the EFT operators are
\begin{eqnarray}
O_1 &=& \bar Q^{(u)} t^a Q^{(u)} \phi^\dagger_4 t^a \phi_3\nn
O_2 &=& \bar Q^{(u)} Q^{(u)} \phi^\dagger_4 \phi_3
\end{eqnarray}
and the EFT anomalous dimension is~\cite{Chiu:2009ft,Chiu:2009mg}

\begin{eqnarray}
\bm{\gamma} &=& \left(2 \gamma_Q + 2 \gamma_\phi\right)\openone +
\bm{\gamma}_S\nn
&& + \frac{\alpha_s}{\pi} \left( -\frac43 i \pi \openone \right)\nn
&&+ \frac{\alpha_2}{\pi}\left(-\frac32 i \pi \openone + \left[ \begin{array}{cc} 
T+U & 2(T-U) \\ \frac38(T-U) & 0 \end{array} \right]\right)\nn
&&+\frac{\alpha_1}{\pi} \left(2 Y_Q Y_\phi (T-U) - i \pi (Y_Q^2+Y_\phi^2)\right)
\end{eqnarray}
where $Y_\phi=1/2$, and
\begin{eqnarray}
\gamma_\phi &=&\left(\frac34 \frac{\alpha_2}{4\pi}+\frac {1} {4} \frac{\alpha_1}{4\pi} \right)\left(2 \log \frac{s}{\mu^2}-4\right)+3 \frac{y_t^2}{16\pi^2}\,.
\end{eqnarray}
The consistency condition Eq.~(\ref{conscond}) is again satisfied by our matching results. Note that the $y_t^2$ term in $\gamma_\phi$ is consistent with the top-quark loop contribution to the matching Eq.~(\ref{tloop}).

\section{Relation between the $S$-Matrix and the Matching Coefficient}\label{sec:smatrix}

The results in this paper are for the on-shell diagrams with dimensional regularization used to regulate the ultraviolet and infrared divergences, and all low-energy scales set to zero. The total amplitude has the form
\begin{eqnarray}
A &=& \alpha \mu^{2\epsilon} T + \alpha^2 \mu^{2 \epsilon} L + \ldots\co
\end{eqnarray}
where $T$ is the tree amplitude, and the one-loop amplitude $L$ contains $1/\epsilon$ UV and IR divergences,
\begin{eqnarray}
L &=& \frac{C_2}{\epsilon^2}+\frac{C_1+D_1}{\epsilon}+\frac{C_2}{\epsilon} \log \frac{\mu^2}{s}\nn
&&+\left(C_1+D_1\right) \log \frac{\mu^2}{s} +\frac12 C_2 \log^2 \frac{\mu^2}{s}+ F(s,t)\fs\nn
\end{eqnarray}
Here $C_{1,2}$ are coefficients of the IR divergences, and $D_1$ is the coefficient of the UV divergence.
Since we have set scaleless integrals to zero, we cannot distinguish IR and UV divergences, and our calculation thus gives $C_1+D_1$, but not each term separately. Note that the coefficient of the $\log \mu^2$ term is proportional to the sum of the $1/\epsilon$ UV plus IR singularitites.

To this must be added the counterterm graphs, which cancel the ultraviolet divergence $D_1/\epsilon$, to give the renormalized $S$-matrix element
\begin{eqnarray}
S &=&  \alpha \mu^{2\epsilon} T + \alpha^2 \mu^{2 \epsilon}\Biggl\{\frac{C_2}{\epsilon^2}+\frac{C_1}{\epsilon}+\frac{C_2}{\epsilon} \log \frac{\mu^2}{s}\nn
	&&+\left(C_1+D_1\right) \log \frac{\mu^2}{s} +\frac12 C_2 \log^2 \frac{\mu^2}{s}+ F(s,t)\Biggr\}\nn
&=& A - \alpha^2 \mu^{2 \epsilon}\frac{D_1}{\epsilon}
\label{93}
\label{smatrix}
\end{eqnarray}
The counterterm graphs must cancel all the UV singularities, since the the theory is renormalizable, so there is no overall $1/\epsilon$ divergence times a $q \bar q VV$ or $q \bar q \phi^\dagger\phi$ operator. Note that the counterterm graphs cancel $D_1/\epsilon$, but not $D_1 \log \mu^2/s$.

The renormalized $S$-matrix has $1/\epsilon$ divergences which are purely IR. These IR divergences lead to IR divergent cross-sections for parton-parton scattering in the massless theory. In QCD, the IR divergences cancel when computing a physical process involving IR safe observables. A textbook example is the cancellation of IR divergences between $e^+ e^- \to q \bar q$ at one-loop, and the tree-level rate for $e^+ e^- \to q \bar q g$, to give a IR safe cross-section for $e^+ e^- \to \text{hadrons}$ at order $\alpha_s$.

 The renormalized $S$-matrix satisfies the renormalization group equation
\begin{eqnarray}
\left[ \mu \frac{\partial}{\partial \mu}+\beta(g,\epsilon) \frac{\partial}{\partial g} \right] S &=& 0
\label{srge}
\end{eqnarray}
where 
\begin{eqnarray}
\beta(g,\epsilon) &=& - \epsilon g - \frac{b_0 g^3}{16\pi^2} + \ldots
\end{eqnarray}
is the $\beta$-function in $4-2\epsilon$ dimensions, with $b_0=11 C_A/3 -2\trwf{}/3-\trcs{} /3$. Applying Eq.~(\ref{srge}) to Eq.~(\ref{smatrix}) shows that
\begin{eqnarray}
D_1 &=& \alpha \frac{b_0}{4\pi}  T\fs
\label{dvalue}
\end{eqnarray}
The one-loop counterterm contribution is equal to the one-loop $\beta$-function times the tree-level amplitude.  Thus we do not need to explicitly compute the counterterm graphs.
Equation~(\ref{dvalue}) and Eq.~(\ref{93}) give
\begin{eqnarray} 
S &=& \alpha \mu^{2\epsilon} T + \alpha^2 \mu^{2 \epsilon} L - \alpha^2 \mu^{2 \epsilon}\frac{b_0}{4\pi} 
\frac{1}{\epsilon} T \ldots\co
\label{expr}
\end{eqnarray}
which relates the renormalized $S$-matrix to the matching condition. An expression analogous to Eq.~(\ref{expr}) can be derived to higher orders. Eq.~(\ref{expr}) is the renormalized $S$-matrix, so all $1/\epsilon$ singularities are IR divergences, which are present in $S$-matrix elements for massless particles.

Using Eq.~(\ref{expr}), we have computed the $q \bar q \to g g$ cross-section and verified that it agrees with Ellis and Sexton~\cite{Ellis:1985er}. Some terms in $A$ do not interfere with the tree amplitude, and hence do not contribute to the cross-section at order $\alpha^3$. The tree-level amplitude only has non-zero helicity amplitudes for $+-$ and $-+$ polarized gauge bosons, so only the real parts of these one-loop helicity amplitudes are checked by the cross-section results of Ref.~\cite{Ellis:1985er}. For example, the matrix element $\mathcal{M}_1$ only contributes to $++$ and $--$ polarization states, and so does not interfere with the tree amplitude. We have also computed the one-loop helicity amplitudes for $++$, $+-$, $-+$ and $++$ polarized gauge bosons from the $S$-matrix Eq.~(\ref{expr}), and verified that they agree with the results in Ref.~\cite{Kunszt:1993sd} for an $SU(N)$ gauge theory. This provides a check on the result of Sec.~\ref{sec:T2d}, which is proportional to $\mathcal{M}_1$.

\section{Conclusion}

We have computed the high-scale matching at one-loop for vector boson production $q \bar q \to V^a_i V^b_j$ and scalar production $q \bar q \to \phi^\dagger \phi$, for an arbitrary gauge theory, and given the group theory factors for the standard model.  When combined with the EFT results of Refs.~\cite{Chiu:2009mg,Chiu:2009ft}, this gives the renormalization group improved amplitudes for gauge boson and Higgs production in the standard model. Numerical plots using these results were already presented in Refs.~\cite{Chiu:2009mg,Chiu:2009ft}. The electroweak corrections to standard model processes at TeV energies are substantial; for example the correction to transverse $W$ pair production at 2~TeV is $37$\%.

\bibliography{matching}

\end{document}